\documentclass[twocolumn]{aastex63}

\newcommand{\degree}{\ensuremath{^\circ}}
\usepackage{csquotes}

\usepackage{natbib}
\bibpunct{(}{)}{;}{a}{,}{,}

\received{\today}
\revised{}
\accepted{}
\shorttitle{Polarization efficiency in Sh II-185}
\shortauthors{Soam et al.}
\graphicspath{{./}{figures/}}
\begin{document}

 \title{Interstellar extinction, polarization, and grain alignment in the Sh 2-185 (IC\,59 and IC\,63) region}

\correspondingauthor{Archana Soam}
\email{asoam@usra.edu}

\author[0000-0002-6386-2906]{Archana Soam}
\affiliation{SOFIA Science Center, NASA Ames Research Center, M.S.
232-12, Moffett Field, CA 94035, USA}

\author[0000-0001-6717-0686]{B-G Andersson}
\affiliation{SOFIA Science Center, NASA Ames Research Center, M.S.
232-12, Moffett Field, CA 94035, USA}

\author{V. Strai\v{z}ys}
 \affiliation{Institute of Theoretical Physics and Astronomy, Vilnius
University, Saul\.{e}tekio al. 3, Vilnius, LT-10257, Lithuania}

\author{Miranda Caputo}
\affiliation{Ritter Astrophysical Research Center, University of Toledo Dept. of Physics and Astronomy, 2801 W. Bancroft St. Toledo, Ohio 43606}
\affiliation{SOFIA Science Center, NASA Ames Research Center, M.S.
232-12, Moffett Field, CA 94035, USA}

\author{A. Kazlauskas}
 \affiliation{Institute of Theoretical Physics and Astronomy, Vilnius
University, Saul\.{e}tekio al. 3, Vilnius, LT-10257, Lithuania}

\author{R. P. Boyle}
\affiliation{Vatican Observatory Research Group, Steward Observatory,
Tucson, AZ 85721, USA}

\author{R. Janusz}
\affiliation{Vatican Observatory, 00120, Vatican City, Italy}

 \author{J. Zdanavi\v{c}ius}
 \affiliation{Institute of Theoretical Physics and Astronomy, Vilnius
University, Saul\.{e}tekio al. 3, Vilnius, LT-10257, Lithuania}

\author{J.A. Acosta-Pulido}
\affiliation{Instituto de Astrof\'isica de Canarias (IAC), C/O Via Lactea, s/n E38205- La Laguna, Tenerife, Spain}
\affiliation{Departamento de Astrof\'isica, Universidad de La Laguna, E-38205 La Laguna, Tenerife, Spain.}

\begin{abstract} Optical and infrared continuum polarization from the interstellar medium is driven by radiative processes aligning the grains with the magnetic field. While a quantitative, predictive theory of Radiative Alignment Torques (RAT) exists and has been
extensively tested, several parameters of the theory remain to be fully
constrained.  In a recent paper, \citet{medan2019} showed that the
polarization efficiency (and therefore grain alignment efficiency) at
different locations in the wall of the Local Bubble (LB) could be
modeled as proportional to the integrated light intensity from the
surrounding stars and OB associations.  Here we probe that relationship
at high radiation field intensities by studying the extinction and
polarization in the two reflection nebulae IC\,59 and IC\,63 in the
Sh 2-185 H II region, illuminated by the B0 IV star $\gamma$ Cassiopeia.  We combine archival visual polarimetry with new 7-band photometry in
the Vilnius system, to derive the polarization efficiency from the
material.  We find that the same linear relationship seen in the Local
Bubble wall also applies to the Sh 2-185 region, strengthening the
conclusion from the earlier study.  \end{abstract}

\keywords{interstellar, dust}

\section{Introduction} \label{sec:intro}

The realization that interstellar polarization of stars is due to
asymmetric dust grains, aligned with the magnetic field, has proven to
be key for mapping the ISM magnetic fields.  Dust-induced interstellar
polarization was first detected by \citet{hall1949} and
\citet{hiltner1949a, hiltner1949b}.  This effect has been used
extensively to derive the magnetic field maps in
different environments \citep[e.g.][]{vrba1976, hodapp1987, bhatt1993,
pereyra2002, matthews2009, chapman2011, sugitani2011}. Polarization of
starlight at visible and near-infrared wavelengths is due to dichroic
extinction by non-spherical dust grains which are aligned with their
short axis parallel to the local magnetic field \citep{whittet2003}.
Aligned dust grains can similarly emit polarized thermal radiation in
the far infrared and at sub(mm) wavelengths \citep[e.g.][]{cudlip1982,
rao1998, dotson2000, vaillancourt2012}.

The mechanism by which the dust grains align with the magnetic field has
long been unclear.  A long assumed mechanism to account for the effect,
based on paramagnetic relaxation in rapidly rotating grain
\citep{davis1951a}, has now been shown to be unviable in most instances
both observationally \citep{hough2008} and theoretically \citep{lazarian1999b}.  However, a radiatively driven mechanism, originally
proposed by \citet{dolginov1976} and more fully developed by
\citet{draine1996, lazarian2007}, has now become the generally assumed
alignment mechanism. This \enquote{Radiative Alignment Torque (RAT) theory} posits that grains are spun up by the transfer of torques from photons in an anisotropic radiation field to helical dust grains.  For paramagnetic grains, the resultant rotation induces a magnetic moment through the Barnett effect \citep{purcell1979}. This induced magnetic moment will cause the grain's angular (and magnetic!) momentum vector to Larmor-precess around an external magnetic field.  Continued RAT torques during the Larmor-precession finally aligns the grain with the magnetic field.  A large number of the predictions of this theory have been confirmed over the last decade \citep[see][for a
review]{bga2015b}, but details of the mechanism and its components
still remain to be probed.

It is important to note that, in most environments, RAT alignment is magnetic alignment; i.e. the grains are aligned with their angular momentum vectors along the magnetic field, and the direction of the observed dichroic extinction polarization traces the plane of the sky projection of the field.  Second order effects -- such as alignment along the radiation field direction, for very strong radiation fields \citep[e.g.] []{2019ApJ...883..122L, 2017ApJ...844L...5K}, and the impact of non-paramagnetic grain materials (carbon solids, e.g. \citealt{2018AAS...23141404A}; Andersson et al. 2020, in prep.) can be important in, rare, extreme environments.
 
While the principles of RAT grain alignment are now well established, we still need to understand the quantitative aspects of the grain alignment
mechanism in context of the strength of radiation and magnetic fields.
\citet{medan2019} presented a detailed study of the interstellar
polarization due to dust in the Local Bubble wall
\citep{lallement2003}, investigating grain alignment and polarization
efficiency dependence on radiation from OB and field stars. They used
polarization measurements from \citet{berdyugin2014} combined with
spectral classifications and photometry from the literature
\citep{wright2003,tycho2000} to investigate the quantitative
relationship between the observed polarization efficiency
($P/A_V$) and the radiation field strength, as well as the strength of
magnetic fields.  In the present work, we perform a similar
investigation of the two nearby (at $d\approx200$~pc)
nebulae IC\,59 and IC\,63 and compare them with Local Bubble results of
\citet{medan2019}.

IC\,59 and IC\,63 are two reflection and emission nebulae in the
Sh 2-185 H\,II region (at $d\approx200$ pc) illuminated by the B0 IV
star, $\gamma$ Cas \citep{karr2005} at a projected distance of $\sim$1.3
pc (IC\,63) and 1.5 pc (IC\,59) from the nebulae (\citealt{bga2013};
Caputo et al. 2020, in press). Sh 2-185 is an H\,II region with a shell of dust with center at $\gamma$ Cas. IC\,59 and IC\,63 are two nebulae in this regions and remain part of this shell.

We have acquired 7-band photometry in the Vilnius system of the
region, which yields both the classification and the visual
extinction of the stars probed.  We analyze these data together with 
existing polarization data from \citet{soam2017} and compare the results to those from the study of the
Local Bubble \citep{medan2019}.

This paper is organized as follows: the polarimetric and
photometric observations are presented in Section \ref{sec:obs}.
Section \ref{sec:res} includes the results of observation acquired for
this work.  Our results are analysed and discussed in Section
\ref{sec:discussion} and we summarize our findings in 
Section \ref{sec:conc}.

\section{Data acquisition} \label{sec:obs}

\subsection{Polarimetry}

\subsubsection{Archival data}


Optical imaging ($R$-band) polarization measurements of nebulae IC\,59
and IC\,63 were also extracted from \citet{soam2017}. These
observations were acquired with the Aries Imaging Polarimeter (AIMPOL)
mounted in the Cassegrain focus of the 104-cm optical telescope,
India.  Details of observations and data reduction can be found in
\citet{soam2017}.

\subsubsection{Observations}

We obtained spectropolarimetric observations from intermediate--dispersion Spectrograph (ISIS) at in the $\sim$f/11 Cassegrain focus of the 4.2 m William Herschel Telescope (WHT) in the Canary Islands, Spain. 
These observations were carried in 2013 on October 18 and 19, as part of the proposal C9-WHT4/13B. The red and blue, both sides were used for acquiring data. We used EEV12 CCD which provides good quantum efficiency down to the atmospheric cut-off and the R300B grating centered at 0.4\,$\mu$m. Red+ CCD and the R158R grating centered at 0.64\,$\mu$m were used for the red arm. We are not discussing blue observations here as those were corrupted by scattered light. The red observations have a coverage of $\lambda \approx 0.48 - 0.95$ $\mu$m. We extracted the polarization values at 0.65 $\mu$m for comparison with archival data. HD\,212311 and HD\,204827 were used as  zero-polarization and high-polarization standards, respectively. The total integration time was divided into 8 positions of half-wave plate (HWP) settings for minimizing the the influence of systematic errors. The standard IRAF routines were used to reduce the data and extract flux. The ordinary ($O$) and extraordinary ($E$) spectra were averaged, for each HWP setting
observation into 0.05\,$\mu$m bins and the polarization calculated by
fitting $(E-O)/(E+O)$ to a cosine function \citep{vaillancourt2020}:
\begin{equation}
 \frac{E-O}{E+O} = a + P*\cos4(\alpha - \theta),
\end{equation}
where $P$ is the amount of polarization, $\alpha$ is the HWP position
angle, 2$\theta$ is the position angle of measured polarization.  The
factor $a$ indicates the gain differences between $O$- and
$E$-beams.  We calculated the normalized Stokes parameters $q (=
Q/I = P\,\cos 4(\theta))$ and $u (= U/I = P\,\sin4(\theta))$.  The
value of $I$ is total intensity Stokes parameter.  The values of
$P$ and $\theta$ are calculated as:
\begin{equation}
    P = \frac{1}{I}\sqrt{Q^{2}+U^{2}}
\end{equation}
and
\begin{equation}
\theta = 0.5\,\arctan(\frac{U}{Q}).
\end{equation}

\subsection{Photometry}

CCD observations in the Vilnius seven-color photometric system of 
two 13$\arcmin$\,$\times$\,13$\arcmin$ areas, centered on the nebulae
IC\, 59 and IC\,63, were obtained with the 1.8\,m VATT telescope on
Mt. Graham, Arizona using the STA0500A camera and a 4k\,$\times$\,4k
CCD chip.  Since the areas had no standards of the Vilnius system,
tie-in observations to the open cluster IC\,1805 were done.
This cluster has Vilnius photometry published by \citet{Straizys2013}.
The angular distance between the IC\,59 and IC\,63 nebulae and the cluster
is about 12 degrees.  Processing of CCD frames has been done with
the IRAF program package in the aperture mode.  The results of
photometry of stars down to $V$\,$\approx$\,19 mag are given in Table 1
(IC\,59, 124 stars; online) and Table 2 (IC\,63, 185 stars; online).  In
these tables the last column gives the spectral and luminosity classes
in the MK system determined from the photometric data with the QCOMPAR
code as described in \citet{Straizys2019}.  The
uncertainties of $V$ magnitudes, and color indices $U$--$V$,
$P$--$V$, $X$--$V$, $Y$--$V$, $Z$--$V$, and $V$--$S$ given in
Tables 1 and 2 take into account the measurement errors and the errors
of transformation to the standard system.

To cover all the stars with the available polarization data, another set
of CCD exposures with a 35$\arcmin$\,$\times$\,35$\arcmin$ field of view,  were obtained with the Maksutov-type 35/50 cm telescope of the
Mol\.etai Observatory in Lithuania, using an Apogee Alta U-47 CCD
camera.  A field center was selected in between the IC\,59 and IC\,63 
nebulae at RA(2000) = 00:58:30, Dec(2000) = +61:01:30.  As in the case of
the VATT observations, a tie-in to the IC 1805 cluster standards was
applied.  The processing of CCD frames and the classification of stars
has been done with the same methods as in the case of the VATT observations,
but the limiting magnitude of Mol\.etai observations is close to $V$ =
17 mag.  The catalog of 487 stars, measured and classified from the
Mol\.etai exposures, is given in Table 3 (online).

The values $A_V$ of the measured stars were determined from their color
excesses $E_{Y-V}$, which are differences between the observed and the
intrinsic color indices $(Y-V)_0$ taken from \citet{Straizys1992}:
\begin{equation}
E_{Y-V} = (Y-V)_{\rm obs} - (Y-V)_0,
\end{equation}
\begin{equation}
A_V = 4.16\,E_{Y-V},
\end{equation}
where the coefficient 4.16 corresponds to the normal extinction law.
Typical uncertainties of $A_V$ due to the observational errors of
$Y$--$V$ and
the errors of the intrinsic $(Y-V)_0$ are $\sim$\,0.1 mag for the stars
with $V < 17$ mag and $\sim$\,0.2 mag for $V$ between  17 and 19 mag.

\begin{figure}
\resizebox{9.0cm}{7.0cm}{\includegraphics{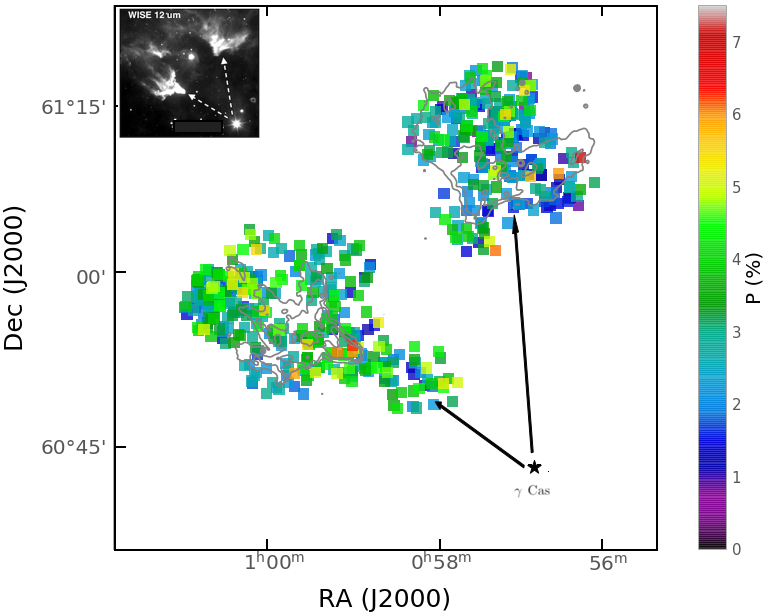}}
\caption{Maps of the amount of polarization observed towards the nebulae
IC\,59 and IC\,63. The inset on left upper corner shows the WISE
12 $\mu$m dust continuum maps of these nebulae. The location of
ionizing star $\gamma$ Cas and the directions of radiation
hitting the nebulae are shown with star symbol and dashed
arrows, respectively.}\label{Fig:pmap}
\end{figure}

\begin{figure*}
\centering
\resizebox{17cm}{9.0cm}{\includegraphics{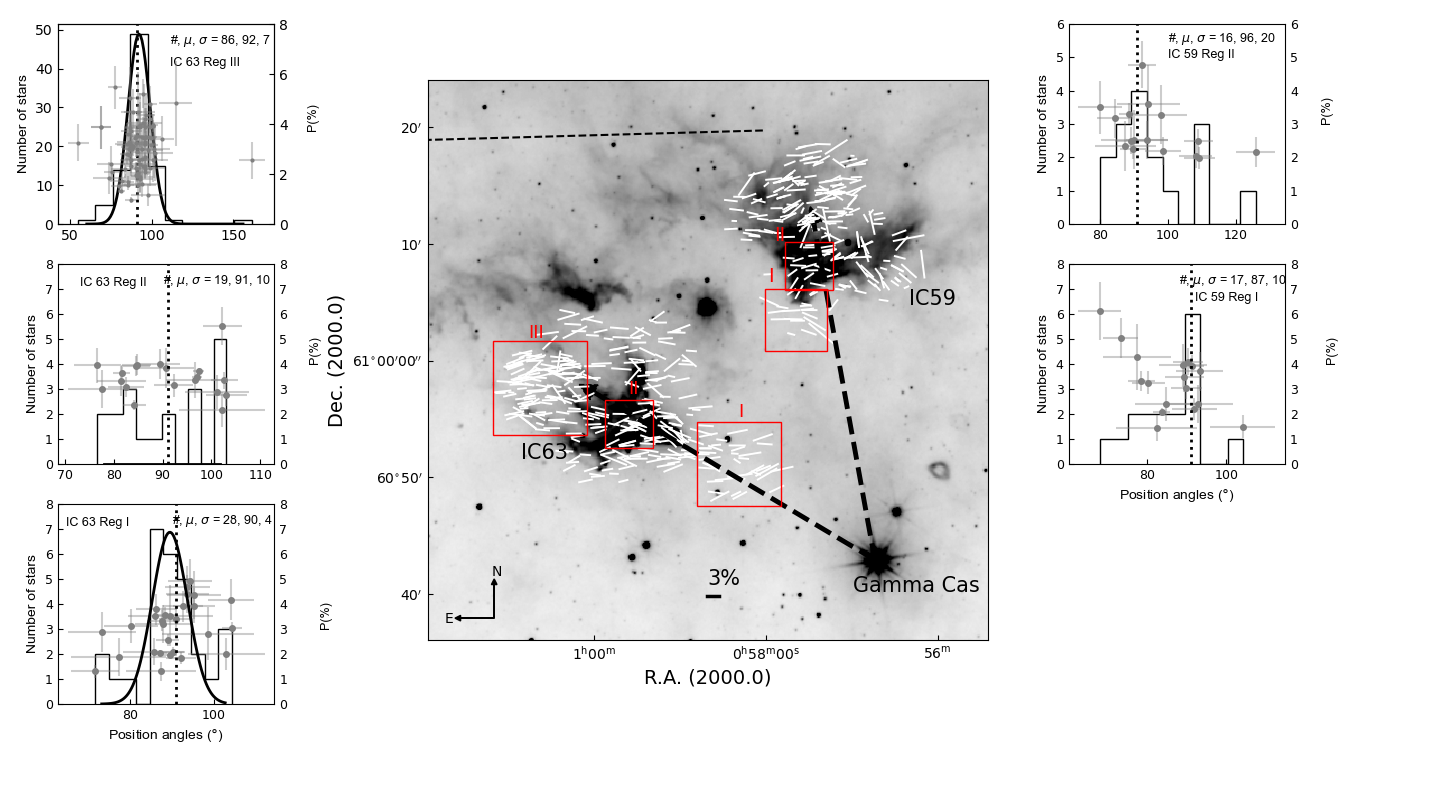}}
\caption{The orientation of the polarization position angles measured
for the stars projected towards IC\,59 and IC\,63 clouds are drawn on
the 48$^{\prime}\times48^{\prime}$ WISE 12\,$\mu$m image.  The nebulae,
IC\,59 and IC\,63, and the ionizing star, $\gamma$ Cas, are identified
and labelled.  The lengths of the line segments correspond to the measured $P$.  A 3\%  line segment is drawn for scaling. 
The arrows with thick broken-line show the direction of the
ionizing radiation.  The arrows pass through the densest part of the
nebulae identified from the 12$\mu$m intensity.  The broken-line shown
on the top left corner represents the inclination of the Galactic plane. The Gaussian
fitted histograms with the distribution of $P$ and $\theta_{P}$
corresponding  to the regions marked in IC\,63 and IC\,59 are also
shown. This figure is adopted from \citet{soam2017}.}\label{Fig:fig1_soam17}
\end{figure*}

\section{Results} \label{sec:res}
Table 4 presents the stars common in polarization measurements of \citet{soam2017} and photometry observations towards IC\,59 and IC\,63.
This table shows columns with the coordinates of stars, polarization values, visual magnitude, extinction, and polarization efficiency values (note that we will use the expression polarization efficiency to denote the observed line-of-sight averaged ratio of polarization and extinction ($\rm P/A_V$)). 

Figure \ref{Fig:pmap} shows the
map of the degree of polarization measured towards the IC\,59 and IC\,63 nebulae in a color scale. The structure of the clouds are
shown with gray solid curves based on the dust continuum emission of
these clouds seen in WISE 12 $\mu$m (inset in upper left corner).  The
location of the ionizing star and the projected directions of the
radiation from $\gamma$ Cas are also shown. A study by Caputo et al. (2020, in press) based on [CII] kinematics found that $\gamma$ Cas lies behind IC\,63 nebula. Therefore, the stars lying outside of the IC\,63 in the plane-of-the sky towards $\gamma$ Cas are not necessarily be significantly close to the $\gamma$ Cas.

\begin{figure}
\resizebox{9.0cm}{7.0cm}{\includegraphics{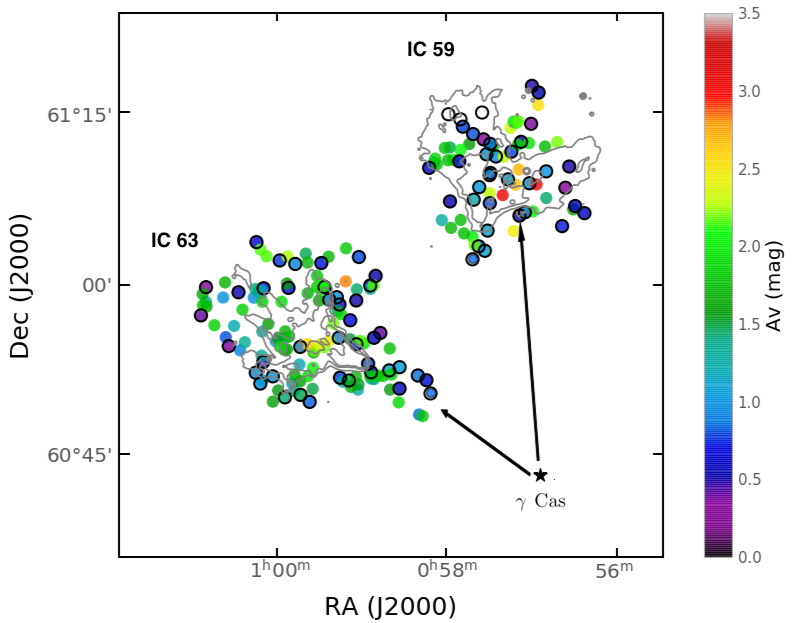}}
\resizebox{9.0cm}{7.0cm}{\includegraphics{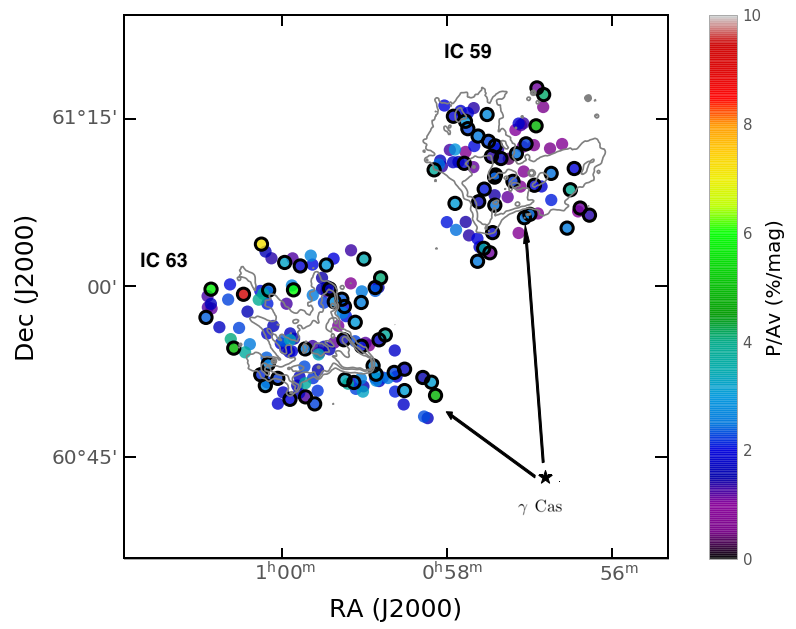}}
\caption{Upper panel shows a map of $A_V$ measurements in IC\,59 and IC\,63. Gray contours are WISE 12\,$\mu$m dust emission. Lower panel is the same as upper panel but for $P/A_V$ values. Symbols with thick boundaries are upto 1.5\,kpc and are used in making Figures \ref{Fig:Av_d} and \ref{Fig:poleff}.}\label{Fig:PAvMaps}
\end{figure}


Figure \ref{Fig:fig1_soam17}, adopted from \citet{soam2017} shows the distribution of the amounts of polarization and the position angles in
different regions (marked) towards IC\,59 and IC\,63 clouds.  This
figure will be discussed in further sections.

The two plots in Figure \ref{Fig:PAvMaps} show the $A_V$ (top) and $P/A_V$ (bottom) maps towards the IC\,59 and IC\,63 nebulae. These $A_V$ and $P/A_V$ values are taken from Table 4 for 78 stars in IC\,59 and 126 stars in IC\,63, respectively. Interstellar extinctions $A_V$
determined from photometry in the Vilnius system are taken from Tables 4. These data are plotted in sky coordinates with their color
indicating the magnitude and \%/magnitude values, as seen in the color
bars, for $A_V$ and $P/A_V$, respectively. The WISE 12 $\mu$m contours in gray color with levels
from $\sim$45\% to $\sim$60\% of the maximum value are plotted to show
the dust structure of the nebulae. The location of $\gamma$ Cas is
shown as a black star, and the black arrows mark the direction of
radiation from this star to the two nebulae.

\begin{figure}
\resizebox{8.5cm}{6.8cm}{\includegraphics{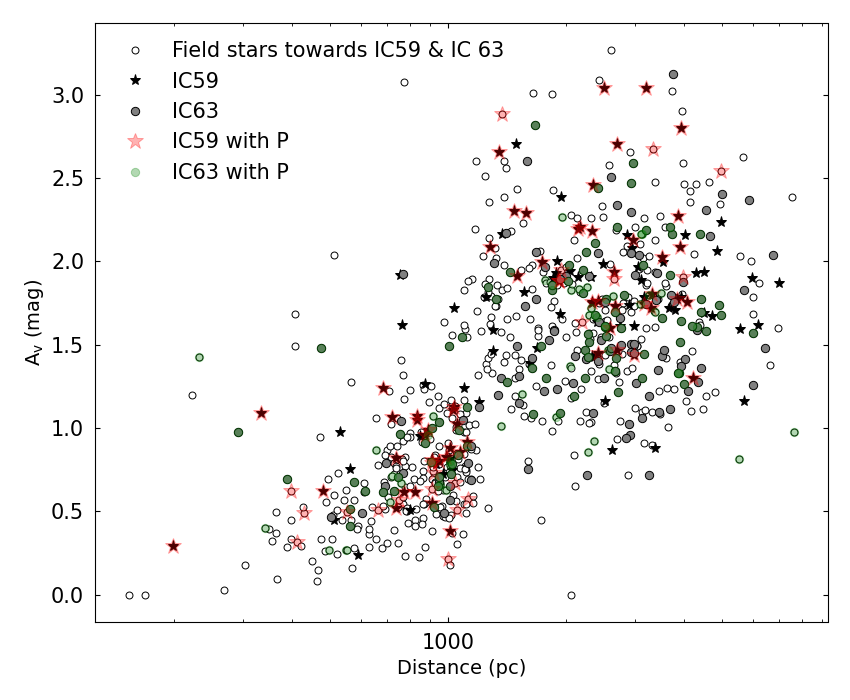}}
\resizebox{8.5cm}{12.0cm}{\includegraphics{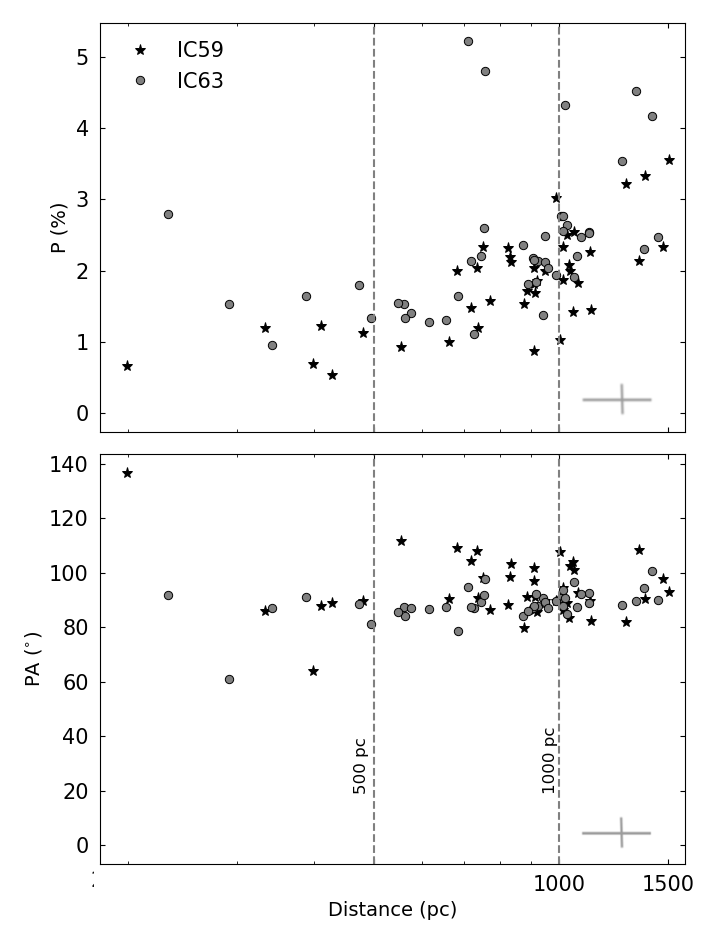}}
\caption{The upper panel shows the distribution of $A_V$ values with
{\it Gaia} DR2 distances from \citet{bailer2018} of all the targets where have photometric information only and colored symbols are those where we have both photonetric and polarimeteric data. The two samples of IC\,59 and IC\,63 are shown with different symbols. The middle and lower panels show the distributions of the degree of polarization and the position angle with distances. Two vertical dashed lines are drawn at distances of 500 and 1000\,pc.}\label{Fig:Av_d}
\end{figure}

The upper panel of Figure \ref{Fig:Av_d} shows the distribution of 
extinction values versus their distances from {\it Gaia} DR2 \citep{bailer2018}. The figure shows an increases in $A_V$ with distances and this distribution becomes more dispersed beyond 1.2\,kpc. The distributions of degree of polarization and position
angles with distances of the targets are shown in the middle and bottom panels of this figure. We limited our analysis to stars with distances of less than 1.5\,kpc. The explanation for this choice is given in section 4.1.

Figure \ref{Fig:cartoon} shows a toy-model/cartoon illustrating the relationship between $A_{V}$ and distance in the nebulae.  This figure is discussed more in Section \ref{sec:layers} for understanding the variations of extinction values with distances of the projected targets on nebulae IC\,59 and IC\,63.

Figure \ref{Fig:lallement14} is reproduced from the original figure adopted from \citet{lallement2014}. More about this figure is given in section 4.1.

Figure \ref{Fig:CO} shows the average results of CO(J=1-0) on-the-fly (OTF)\footnote{In this mode of observations, data is acquired with telescope pointing moves between two points on the sky. The telescope slews in the given stripe duration at a constant speed linearly in RA and linearly in Dec. The OTF mode is typically used to cover large sky areas in short time.} observations towards IC\,63 (Soam et al., in prep.). The most prominent emission lines are seen at radial velocities of --20 and 0\,km/s. 

Figure \ref{Fig:poleff} shows the polarization efficiency (i.e. $P/A_V$) values for background stars with measured extinctions in the directions of the IC\,59 and IC\,63 nebulae.  The best fitting lines to the distributions are also plotted along with equations containing the best-fit parameters in each plot.  For plotting these distributions, we have considered only those targets which are at or less than 1.5\,kpc away.

The polarization efficiency ($P_V/A_V$) variation with extinctions for LDN\,204- Cloud 3 is shown in Figure \ref{Fig:L204} by using archival data from \citet{cashman2014}.

Figure \ref{Fig:comp_bubb} shows the distribution of polarization efficiencies as a function of radiation flux from the illuminating sources.  This plot presents a comparison of the Local Bubble, with UV flux from OB associations \citep{medan2019} with our findings towards IC\,59 and IC\,63 nebulae affected by the UV flux from $\gamma$\,Cas. We also plotted values for LDN\,204 adopted from \citet{cashman2014}.

A cartoon shown in Figure \ref{Fig:psi_fig} is adopted from \citet{medan2019} to illustrate the inclination angle ($\psi$) between the line-of-sign vector and the cloud region.

\section{Analysis and discussion}\label{sec:discussion}

\subsection{Polarization efficiency and extinction at different
distances}\label{sec:layers}

To understand the effect of illumination on the grain alignment, we must isolate the dust in, and therefore polarization from, the IC\,59 and IC\,63 nebula from possible background polarization.  We can do this based on the extinction versus distance plots.  There are two, likely complementary, origins of steps in these plots (Figure \ref{Fig:Av_d}). If the surface density (i.e. localized $A_{V}$) of each individual extinction layer is uniform, then each step in $A_{V}$ versus $d$ represents a new layer of dust extinction at the
distance of the step.  However, multiple steps in $A_V$ versus $d$ can
also occur with a single physical, but non-uniform, layer.

We can see this by considering a uniform, but relatively sparse, space
distribution of background stars above a given apparent magnitude (i.e.
such that they are included in a magnitude limited photometric survey).
The surface density on the sky of observed stars will be the
product of their space density and the distance to
which the observations extend.  The surface density of measured stars
will then rise linearly with distance probed, or equivalently, the
on-the-sky distance between observed stars will shrink with increasing
line of sight distance sampled.

\begin{figure*}
\centering
\resizebox{13.0cm}{7.0cm}{\includegraphics{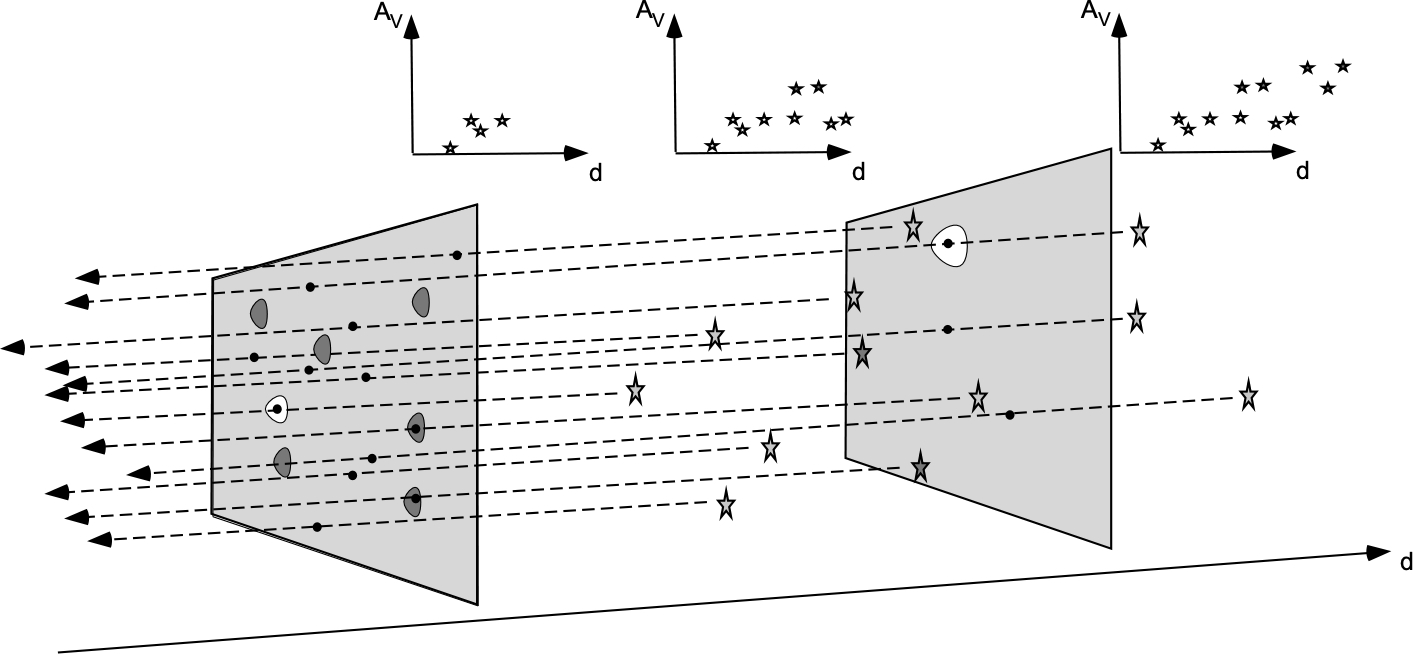}}
\caption{The two causes of extinction steps in an extinction vs. distance plot are illustrated in a toy-model/cartoon. For a cloud with a density structure -- especially where the surface area of the high extinction regions is significantly smaller than the average extinction of the cloud/layer -- an extinction jump may be detected at a distance corresponding to where the chance of a line of sight encountering a \enquote{clump} becomes significant.  The lower envelope of the distribution will, however stay fixed. When a second extinction layer is encountered both the upper and lower envelopes of the distribution will change. For extinction layers with low density sub-regions (\enquote{holes} shown as white regions in the screens), a small number of low A$_V$ outliers may be expected.  These different behaviors can all be seen in the top panel of Figure \ref{Fig:Av_d}.}\label{Fig:cartoon}
\end{figure*}

If an extinction layer has density enhancements that are small on the
scale of the total area surveyed (and, again, the stellar surface
density is limited), the probability that such a dense sub-region
(Figure \ref{Fig:cartoon}) intercepts the line of sight to a background
star will then be significantly less than one.  If the extinction
density enhancements have a characteristic scale size, then an
\enquote{extinction step} will occur at the distance where that scale
size is first effectively sampled by the stellar surface density.  For
a spectrum of extinction enhancement scales, we would expect a more
distributed increase in $A_{V}$ versus $d$, yielding increased dispersion
in extinction for a given stellar distance. 

In this scenario of an extended low-extinction layer with embedded
localized higher extinction regions, we would expect the upper envelope
of the extinction to rise either abruptly -- for a single scale
size of denser regions -- or gradually for a distribution of high-$A_V$
\enquote{clumps}.  The lower envelope, corresponding to the large scale
low-extinction part of the layer should, however, stay close to
constant. For an additional distinct extinction layer, the lower
envelope of the $A_V$ distribution should also rise. 

It is, of course, also possible to consider an interstellar
density distribution where the dense (higher $A_{V}$) regions of a
physical extinction layer are more broadly distributed with smaller, low
$A_V$, \enquote{hole} areas interspersed (see Figure \ref{Fig:cartoon}). This would, however, still raise both the upper and lower envelopes of the $A_V$ distribution at
the characteristic distance, with a small number of lower $A_V$ points
remaining, and should still be differentiable from a uniform physical
layer or a single inhomogeneous one. This is illustrated in the distributions of $A_V$ vs. d panels of Figure \ref{Fig:Av_d}.

Therefore, the better diagnostic of a true additional extinction layer
with distance will be the minimum extinction as a function of the
distance, rather than the maximum extinction as a function of the
distance. The upper panel of Figure \ref{Fig:Av_d} illustrates this for
the direction of Sh 2-185.  As is clear from the figure, while there
are enhancements of the maximum extinction towards IC\,59 and
IC\,63 between 700\,pc$-$1.2\,kpc, the lower envelope of the
extinction remains approximately constant all the way from $\sim$200\,pc
to $\sim$1.2\,kpc.  At about 1.2--1.5\,kpc a significant increase occurs
in both the upper and lower envelopes of the extinction distribution,
with a small number of low extinction points remaining at
$A_V\lessapprox$1 mag. 

This interpretation is consistent with the trigonometric parallax
distances to $\gamma$ Cas, on the one hand, and the Perseus arm,
on the other.  The {\it Hipparcos} distance for $\gamma$\,Cas
(\textit{l,b}=124, -2) is $188\pm20$\,pc.  The distance to the Perseus
arm can be estimated from the {\it Gaia} distance to the OB stars in the
star forming complex W3 (\textit{l,b}=134, +1) of $\approx$\,2.2\,kpc
\citep{2019MNRAS.487.2771N}.  Since the OB stars are expected to be
concentrated towards the center of the spiral arm, we could expect the dust in the arm to be detectable somewhat closer than the OB stars.  We therefore assume that our
subsequent analysis that stars out to 1.5\,kpc probe only the extinction
from dust in Sh\,2-185. 

\begin{figure*}
\centering
\resizebox{14cm}{11cm}{\includegraphics{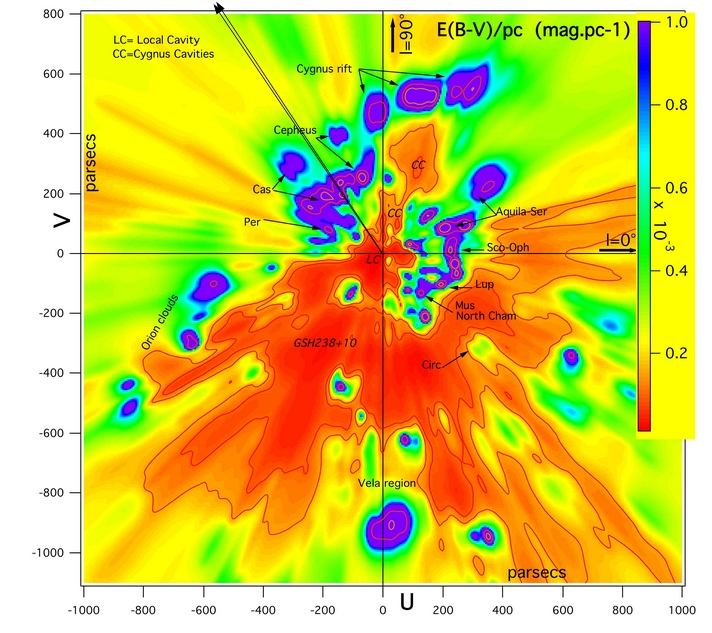}}
\caption{This figure adopted from \citet{lallement2014} and reproduced with permission \textcopyright ESO. The inverted differential opacity distribution in the Galactic plane in the Solar neighborhood is shown. The sun is located at (0,0) coordinates with the Galactic center toward the right. The color scale from red to violet shows increasing differential opacities. The two arrows are scaled to be 1 kpc in length, and drawn to bracket the extent of the Sh 2--185 region (i.e. from just west of IC\,59 to just east of IC\,63).  The extinction distribution in this map supports our conclusion of very little foreground extinction, or background extinction to at least 1 kpc, in the direction of Sh 2-185.}\label{Fig:lallement14}
\end{figure*}

Figure \ref{Fig:lallement14}, adopted from \citet{lallement2014} shows the differential opacity in the Galactic plane out to $\sim$1\,kpc.  The two over-laid arrows are scaled to represent 1\,kpc distance and span the Galactic longitude range of 123.45$\degree$ to 124.05$\degree$ (i.e. from just west of IC\,59 to just east of IC\,63).  As the figure shows, the differential opacity in this direction is low out to $\sim$200\,pc where a high-opacity ridge intercepts our lines of sight (as Figure 3 in Lallement et al. shows, the sampling density of their data indicates that the depth of this ridge is likely, at best, marginally resolved).  Beyond this ridge no further high-opacity material is seen to the extent of the mapping (although beyond $\sim$700\,pc the spatial resolution in the Lallement et al. data decreases, and therefore small-scale structure may not be detected close to the Perseus arm).  Thus, our interpretation of our $\rm A_{V}$ vs. distance plots is consistent with the extinction mapping of \citet{lallement2014}.

\begin{figure}
\resizebox{9.0cm}{5.8cm}{\includegraphics{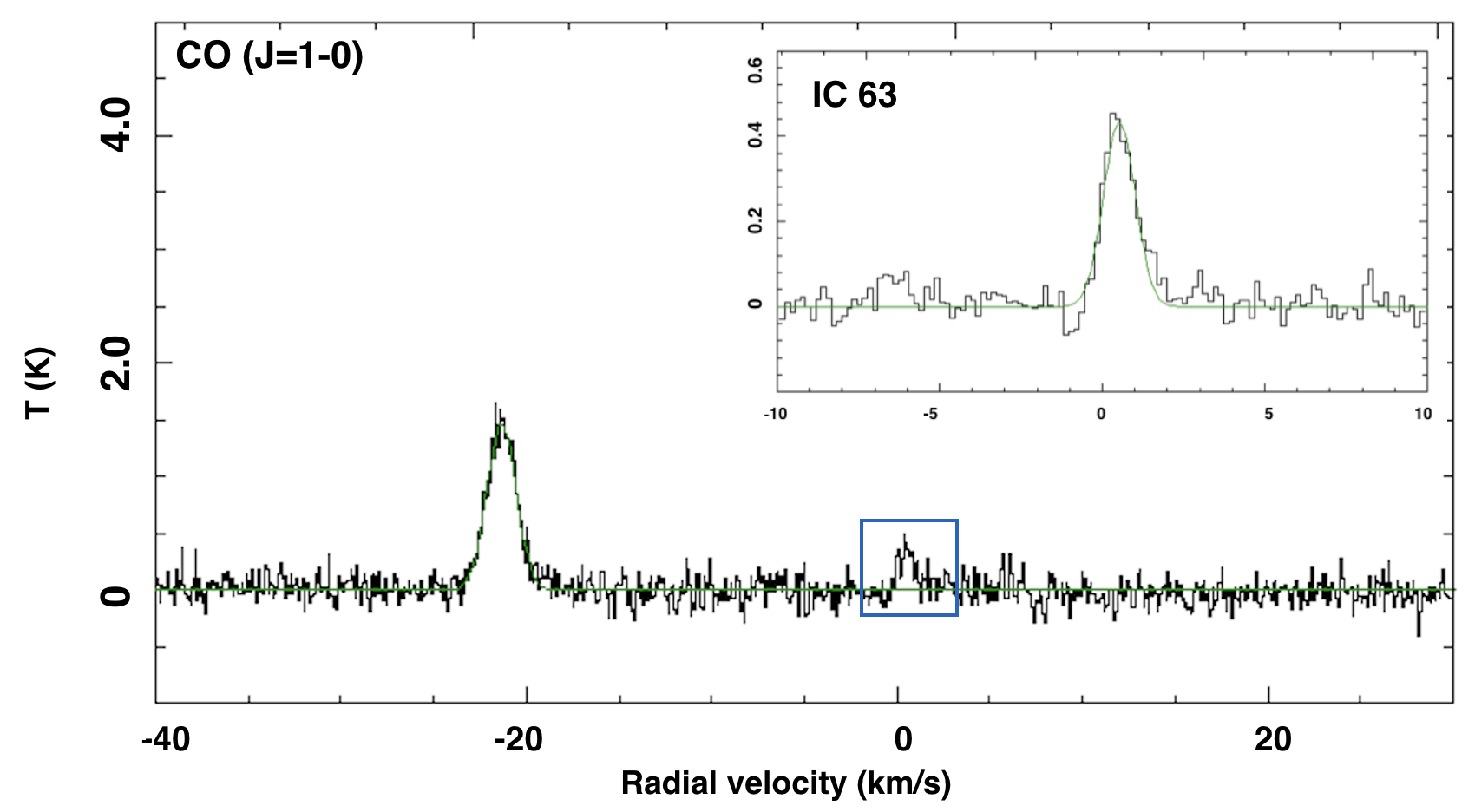}}
\caption{Emission towards IC\,63 in CO(J=1-0) line (Soam et al., in prep.) The dominant peaks are seen at radial velocities close to --20 and 0\,km/s. The line feature at 0\,km/s framed with blue box is zoomed-in at the upper right corner of the figure.}\label{Fig:CO}
\end{figure}

In addition, we investigated CO (J= 1-0) data from \citet{dame2001} at multiple positions in the near vicinity of IC\,59 and IC\,63. All positions show a strong CO band with radial velocity close to --20\,km/s.  According to the \enquote{Kinematic Distance Calculator} \citep{reid2009}, this velocity gives a distance of the cloud close to 1.5\,kpc. This is in agreement with our $A_V$ vs. distance plots for the VATT photometric data (upper panel in Figure \ref{Fig:Av_d}) which at about 1.5\,kpc show a steep rise of the extinction up to Av close to 3 magnitudes. It is interesting that the Dame et al. radio data do not show any significant intensity maximum at radial velocity close to zero which would correspond to the IC\,59 and IC\,63 clouds at 200 pc. We note that the \citet{dame2001} surveys are sampled only on a 7.5\arcmin\ grid and hence the small molecular cores of IC\, 59 and IC\, 63 would be significantly beam-diluted in their data. However, we found some significant CO(J=1-0) emission at radial velocity close to zero towards IC\,59 and IC\,63 in our 13-m single dish observations (Soam et al., in prep.)\footnote{A detailed analysis of line observations is beyond the purpose of this paper and will be presented in a work under progress.} from Taeduk Radio Astronomy Observatory (TRAO) with much better resolution ($44\arcsec$) than that of observations of \citet{dame2001}. Figure \ref{Fig:CO} shows the emission at 0\,km/s and close to --20\,km/s radial velocities. \citet{jansen1994} also reported a detection of CO (J=2-1) in IC\,63 nebula at radial velocity ~0.6\,km/s. \citet{heyer1998ApJS} found a CO (J=1-0) emission at velocity close to zero km/s located behind the curved surface of IC\,59 facing radiation from $\gamma$\,Cas. Our recent observations of CO (J=1-0) and these previous studies show that there is a prominent CO emission towards IC\,59 and IC\,63 nebulae. As we discussed above on emission from six positions around these nebulae from Dame et al., the emission seen at --20\,km/s might correspond to Perseus arm. From these arguments, we conclude that no strong widespread absorbing and polarizing clouds are present behind $\gamma$\,Cas up to 1 or 1.5\,kpc.



\subsection{Grain alignment due to external radiation and comparison with Local Bubble}\label{sec:beta_calc}

\begin{figure}
\resizebox{9.0cm}{7.6cm}{\includegraphics{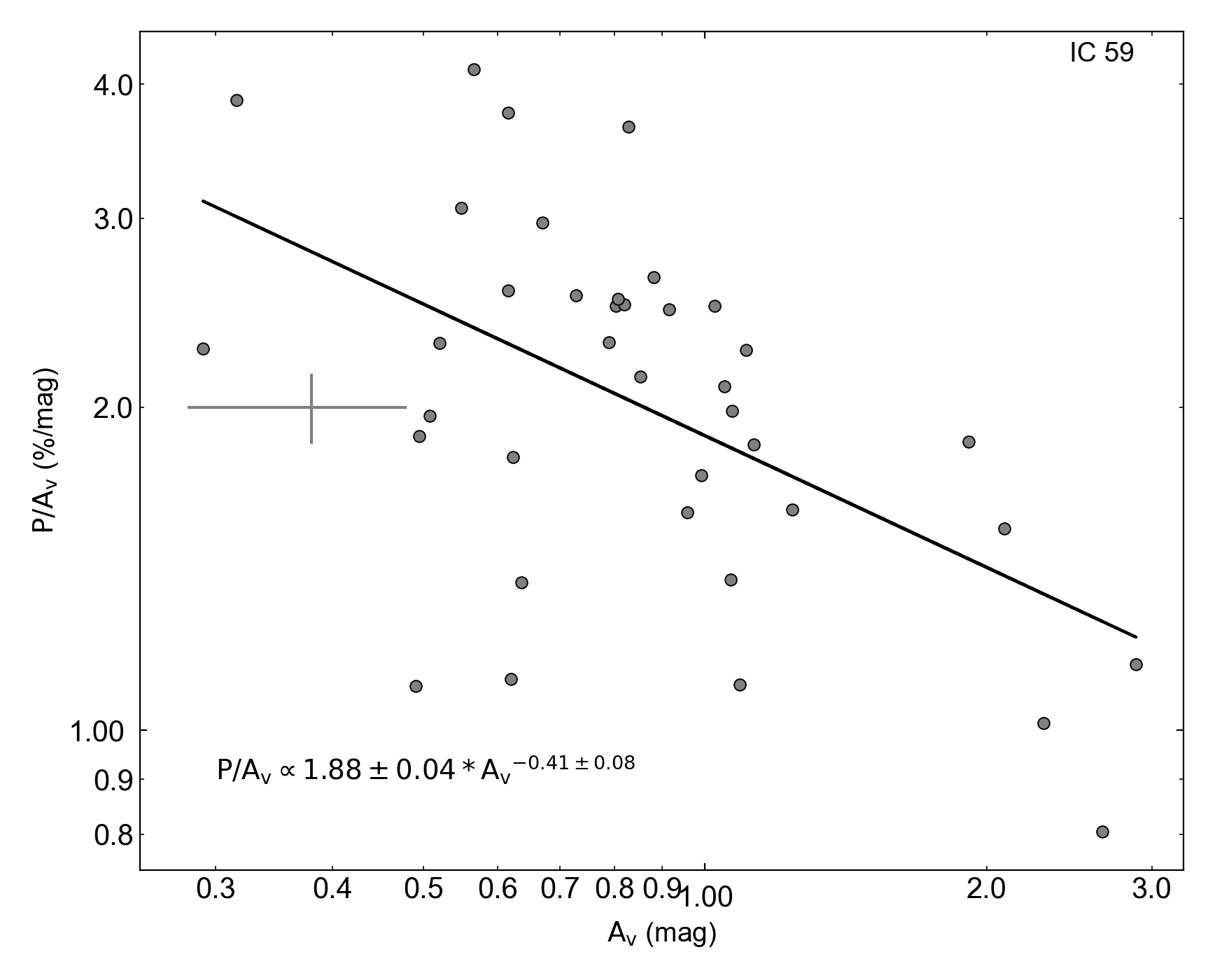}}
\resizebox{9.0cm}{7.6cm}{\includegraphics{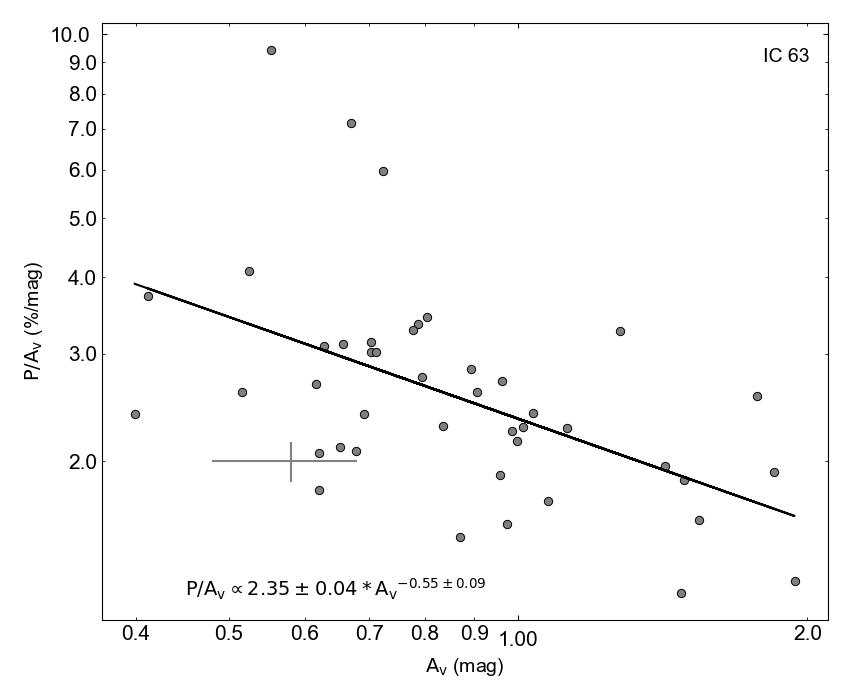}}
\caption{The variation of polarization efficiencies with extinction
values measured towards the targets projected on the IC\,59
(upper) and IC\,63 (lower) nebulae. The best fitted parameters and
typical error bars are also indicated.}\label{Fig:poleff}
\end{figure}

\citet{cashman2014} studied the variations of H--band
polarization efficiency ($P_H/A_V$) in the LDN\,204- Cloud 3 with distance from
$\zeta$ Oph (spectral class O9.5\,V).  They found that the polarization
efficiency steadily decreases with the distance from $\zeta$ Oph.  Also
the power law fit to the distribution of polarization efficiency and
extinctions shows a steep index of --0.74$\pm$0.07 which suggests less
grain alignment at high extinctions compared to other studies performed
on different molecular clouds. They noticed (beyond the uncertainties)
that the weighted mean polarization efficiencies ($P_{H}/A_V$) do vary
systematically with distance to $\zeta$ Oph.

\citet{medan2019} performed a more general study of the grain alignment
variation in the Local Bubble wall, using a large polarization survey of
the North Galactic cap from \citet{berdyugin2014}.  They found that the
polarization efficiency is linearly dependent on the intensity of the
illumination of the grains, and is dominated by the light from the OB
association within 150\,pc. Adding in contributions from field stars
also allowed them to investigate the wavelength dependence of RAT
alignment, and to show that the alignment is most sensitive to the blue
light from O and B type stars.

In the present study, we add a well defined system with two nebulae 
radiated upon by higher radiation flux from $\gamma$ Cas, a B0\,IV star, in the Sh 2-185 H\,II region.

For some regions in Figure \ref{Fig:pmap} no projected star was bright enough to yield significant polarization detections. The spatial distribution
of polarization efficiencies shown in lower panel of Figure \ref{Fig:PAvMaps} indicates relatively higher polarization efficiencies on the cloud boundaries in low density regimes.  Figure \ref{Fig:fig1_soam17}, adopted from
\citet{soam2017}, shows a similar trend, where the polarization
measurements in regions I, II, and III in IC\,63 and regions I and II in
IC\,59 [labelled in red color], show a polarization fraction which is relatively higher in regions closer to $\gamma$ Cas, especially in IC\,59. 
The polarization efficiency as a function of visual extinction can be
interpreted in the framework of \citet{jones1992}:

\begin{figure}
\centering
\resizebox{9.3cm}{7.2cm}{\includegraphics{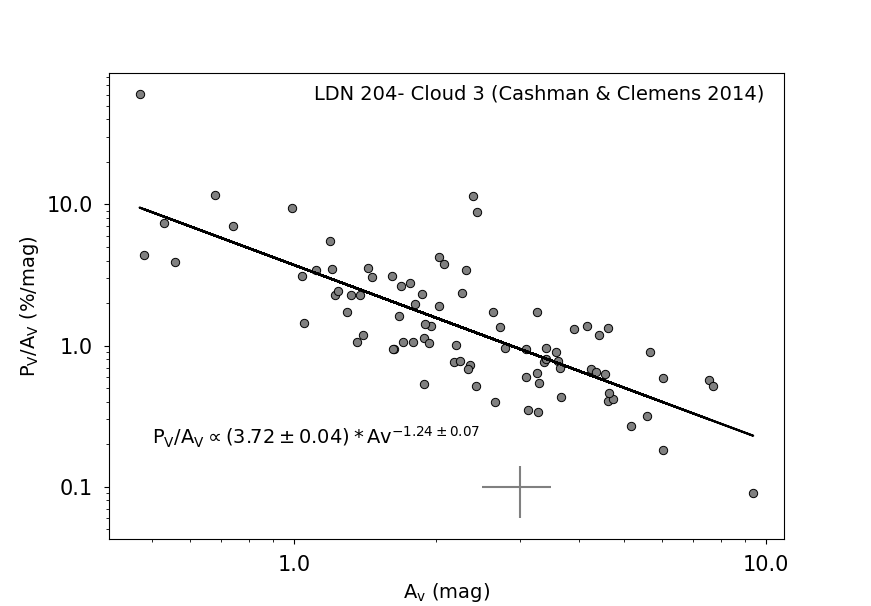}}
\caption{ Polarization efficiency ($P_{V}/A_{V}$) vs. extinction distribution of stars observed towards LDN 204-Cloud 3. The data is adopted from \citet{cashman2014}. From their archival data, we used only those detections where $P/\sigma{P} > 3$ is satisfied. The H-band polarization values are transformed to V-band using the relation from \citet{bga2007}.}\label{Fig:L204}
\end{figure}

\begin{equation}\label{eq:polAv}
\frac{P}{A_{V}} = \beta {A_{V}}^{\alpha},
\end{equation}
where $\alpha$ depends primarily on the turbulence of the medium and --
for large extinctions -- grain alignment variation \citep{alves2014,
Jones2015}.  For a fully turbulent medium, $\alpha$ is expected to take
a value of --0.5, indicating a random walk through a large number of
turbulent cells with differently oriented magnetic fields
\citep{jones1992}.  The value at $A_V=1$ can be used as a comparison of
the polarization efficiency between samples.  Under the assumption that the inherent
dust characteristics (roundness, size distribution, mineralogy etc.) are
similar for the different regions, $P/A_V$ at $A_V=1$ can be
used as a tracer of grain alignment efficiency.

\begin{figure*}
\centering
\resizebox{17.0cm}{12.0cm}{\includegraphics{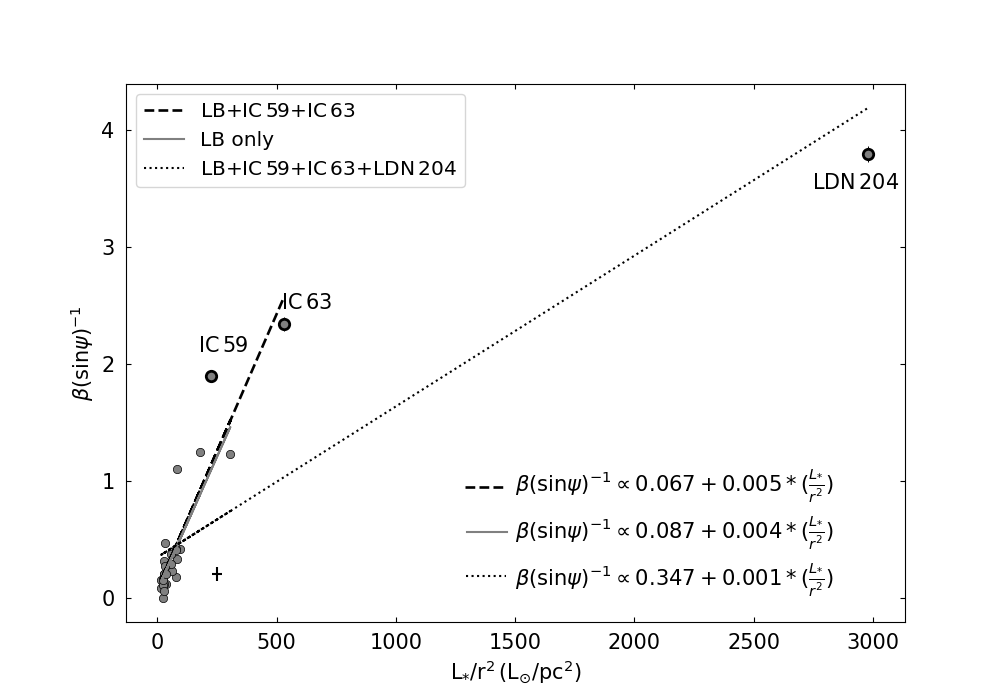}}
\caption{The variation of the polarization efficiency indicated by
$\beta(\sin\psi)^{-1}$ \citep[see][]{medan2019} with UV photon flux. The values measured toward the Local Bubble \citep{medan2019} are shown with gray filled circles and those of IC\,59, IC\,63 [this work] and LDN\,204 \citep{cashman2014} with errors are shown as gray filled circles with thick black boundaries. Typical error bar on Local Bubble measurements is also shown. The source of UV fluxes towards IC\,59 and IC\,63 nebulae is $\gamma$\,Cas and for LDN\,204, it is $\zeta$ Oph. The walls of local bubble receives UV flux from OB associations.}\label{Fig:comp_bubb}
\end{figure*}


We used Eq. \ref{eq:polAv} to fit the polarization and extinction
measurements towards IC\,59 and IC\,63 shown in Figure \ref{Fig:poleff}. We considered data upto 1.5\,kpc only for plotting and fitting in this figure. The reason for choosing this distance is given in section 4.1. We find values of $\alpha$ and $\beta$ of --0.55$\pm$0.09 and 2.35$\pm$0.04 for IC\,63 and --0.41$\pm$0.08 and 1.88$\pm$0.04 for IC\,59, respectively. As we stated above in section 2, we used a typical error of 0.1\,mag in $A_V$ for calculating uncertainties in fitted values. To evaluate the sensitivity of the results to systematic errors (such as the incorrect assignment of any polarization or extinction to the two clouds, we systematically removed one line of sight from the sample for each cloud and recalculated the power-law fits.  The resulting sample of fit parameters are consistent with the fit to the full sample for each cloud, indicating that the fit results are dominated by random errors and that the results are robust.  For IC\, 63 the $\alpha$ parameter distribution has a mean of -0.55 with an error on the mean of 0.03, while the $\beta$ parameter has a mean of 2.34 with an error on the mean of 0.02. For IC\, 59 the two distributions have means and error on the means of $\alpha$:(-0.45,0.03) and $\beta$:(1.89,0.07), respectively.

\citet{cashman2014}, in their study of Cloud\,3 of LDN\,204 near
$\zeta$\,Oph, found a slope in P$_H$/A$_V$ of $\alpha=-0.74\pm0.07$. Because of the shape of the Serkowski curve \citep{serkowski1973}, the NIR polarization is significantly less than that for optical bands, for almost any extinction. 
We therefore need to convert the H-band data from \citet{cashman2014} to estimated V-band polarization, in order to compare their results to the Local Bubble wall results. In addition, as suggested by \citet{whittet2001} and confirmed and expanded on by \citet{bga2007} a universal relationship exists between $\lambda_{\rm max}$ and $A_V$, with a
secondary dependence on the average total--to--selective extinction
($R_V$) in each region. Because the peak of the polarization curve shifts to longer wavelength with increasing extinction, the transformation from P$_H$ to P$_V$ will depend on the extinction of the line of sight, and will result in a steeper P$_V$/A$_V$ vs. A$_V$ relation than for P$_H$/A$_V$ vs. A$_V$. We used the $\lambda_{\rm max}$ vs. $A_V$ relationship from \citet{bga2007} together with extinction estimates
from \citet{cashman2014} to convert the H--band measurements to estimates
of the V--band polarization, and thence derive $P_V$/$A_V$ values. Performing this conversion from P$_H$/A$_V$ to P$_V$/A$_V$ yields an equivalent, V-band, $\alpha$ value for LDN\,204 of $\alpha=-1.24\pm0.07$.

Figure \ref{Fig:comp_bubb} shows the variation of polarization efficiency represented by  $\beta(sin\psi)^{-1}$ (assuming $\sin(\psi)$=1 i.e. $\psi=90^\circ$, $\beta$ is equivalent to a lower limit of the alignment efficiency (P/A$_V(A_V$=1), Eq. \ref{eq:polAv}) with flux ($L_{*}/r^{2}$, where $\rm L_{*}$ is the luminosity of $\gamma$ Cas). These expressions are explained later in this section.

As is shown in Figure \ref{Fig:comp_bubb}, the values of polarization
efficiencies ($P/A_V$ ($A_V=1$)) are significantly higher in Sh 2-185
than those found in the Local Bubble wall with $\beta=1.95\pm 0.05$ for IC\, 59 and $\beta=2.24\pm 0.04$ for IC\, 63. This indicates a higher
grain alignment efficiency in the Sh 2-185 region. Similarly to the
results by \citet{cashman2014} in the LDN\,204 cloud, we also derived a
decrease in polarization efficiency for IC\,63 and IC\,59 with
increasing distance from $\gamma$\,Cas. The polarization efficiency in this context refer to $\beta(sin\psi)^{-1}$ which is inversely proportional to the square of the distance (r) of the cloud from $\gamma$ Cas. As seen in Figure \ref{Fig:comp_bubb}, $\beta(sin\psi)^{-1}$ value of IC\,59 is less than that of IC\,63. This is because IC\,63 is a relatively closer (r=1.3\,pc) to $\gamma$ Cas than IC\,59 (r=1.5\,pc). For LDN 204 using the P$_V$/A$_V$ plot we derive $\beta=3.72\pm0.04$.

\begin{figure}
\resizebox{7.5cm}{7cm}{\includegraphics{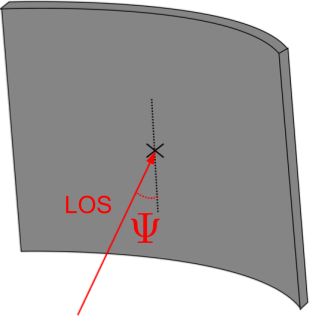}}
\caption{Adapted from \citet{medan2019}. The diagram represents one of the regions of LB wall. The center of the region is intersected by line-of-sight vector in red color. The inclination angle between the line-of-sight vector and the shaded cloud region is represented by $\psi$.}\label{Fig:psi_fig}
\end{figure}

To test the variation of radiatively driven grain alignment with
distance of the cloud from the radiation sources, \citet{medan2019} constructed a model 
to predict values of $\beta(sin \psi)^{-1}$ for the Local bubble. The model was defined as:

\begin{equation}
\beta(\sin\psi)^{-1} (\ell, b) \propto \frac{L_{*}}{(x_{*}-x_{i})^{2}+(y_{*}-y_{i})^{2}+(z_{*}-z_{i})^{2}}
\end{equation}

where $L^{(i)}_{*}$ is luminosity of the source $i$, $x_{*}$ (etc.) are the locations of the source, and $x_{i}$ (etc.) refer to the position at the nebula [wall of LB]. Combining the all sources in LB, the model of \citet{medan2019} takes the form: 

\begin{equation}\label{eq:il1}
\beta(\sin\psi)^{-1} = A + B\cdot\Sigma\frac{L^{(i)}_*}{r^2_i}
\end{equation}

where $L^{(i)}_{*}$ is the luminosity of the source $i$ at a distance $r_i$ from the cloud. The function $\sin\psi$ accounts for  the fact that dust-induced polarization only probes the magnetic field component in the plane of the sky.  Since the grains are spinning around the field lines, no polarization is generated along the magnetic field direction \citep[see][for details]{medan2019}. Figure \ref{Fig:psi_fig} illustrates the concept.

Under the assumption of RAT alignment the alignment efficiency will depend on the intensity of the radiation field at the location of the grain.  This will, in turn, depend on the extinction between the illumination source and the grain, and, if the radiation field is dominated by a point source, the distance from the source to the grain (through the r$^{-2}$ dependence).  For regions with rapidly varying radiation fields (i.e. close to a point source) or with significant extinction, the measured polarization efficiency (P/A$_V$) may sample regions with different alignment efficiency, especially at higher observational line-of-light visual extinction.  However, because of the RAT condition, $\lambda<2\textit{a}$, and because the extinction curve falls to the red, RAT alignment is relatively insensitive to the extinction between the source and the grain for moderate column densities, which will somewhat lessen the variability relative to the visual ($\lambda$=0.55 $\mu$m) extinction.  In addition, along an observed line of sight the polarization efficiency will probe dust and gas at varying levels of source-to-grain extinction in a poorly constrained way that depends on the density and location of the cloud relative to the illuminating source.  Because the power-law fits in \citet{medan2019} and Figure \ref{Fig:poleff} (this work) are dominated by line-of-sight extinction with A$_V<$1 mag., internal radiation damping in these clouds should not have a significant impact on the fitted $\alpha$ or $\beta$ values.  In Figure \ref{Fig:L204}, which shows the data from \citet{cashman2014} after transformation from P$_H/A_V$ to P$_V/A_V$ (from their archival data, we used only those detections where $P/\sigma{P} > 3$ is satisfied. The H-band polarization values are transformed to V-band using the relation from \citet{bga2007}), the larger best-fit $\alpha$ value may indicate a significant contribution of cloud-internal extinction vis-\`a-vis $\zeta$\, Oph (suppressing the alignment efficiency, and therefore P/$A_V$, at the higher extinctions).  However, as the points at A$_V<1$ are continuous with the ones at A$_V>1$, the $\beta$ value is unlikely to be severely affected, certainly not to the level of $\beta \approx$13 that the straight line of the Local Bubble wall and Sh 2-185 results would imply\footnote{if we calculate $\beta (sin\psi)^{-1}$ using best fitted A and B values of Local bubble and IC\,59 and IC\,63 nebulae and $L/r^{2}\approx$ 3000 of LDN\,204, we get a value close to 13.}.

For the region between Galactic latitude 30--42$^\circ$, \citet{medan2019} found $A=0.0055\pm0.0103$ and $B=0.0047\pm0.0002$ from the best fit of model predicted radiative driven grain alignment based on luminosities of local OB association versus the measured grain alignment (see Figure 12 of \citealt{medan2019}).

For IC\,63 and IC\,59 we assume that the magnetic field is oriented close to the plane of the sky and, therefore $\sin(\psi)$ is unity.
\citet{crutcher2003} showed that the average Galactic magnetic field in the Solar neighborhood is directed towards \textit{l,b}$\approx80.6,0$.  Since Sh 2-185 is located at \textit{l}$\approx 124^\circ$ this means that the Galactic magnetic field, nominally, makes an angle of $\psi\approx46^\circ$ with the clouds in the region.  However, because of the dynamics of a magnetized H II region \citep{stil2009,gendelev2012}, it is not clear that the magnetic field threading IC 59 and IC 63 agree with the direction of the Galactic average.  As shown by the polarization mapping by \citet{soam2017} a significant plane-of-the-sky component of the magnetic field is present in the nebulae. As noted above, an assumption of $\sin(\psi)$=1 (i.e. $\psi=90^\circ$) makes $\beta$ equivalent to a lower limit of the alignment efficiency (P/A$_V(A_V$=1), Eq. \ref{eq:polAv}).
Given the distance of the nebulae from $\gamma$ Cas of 1.3 and 1.5\,pc
\citep{Hoang2015}, we can put the results for Sh 2-185 on the same plot
as those for the Local Bubble wall.

Figure \ref{Fig:comp_bubb} shows the values of polarization efficiency
modified by $\beta(\sin\psi)^{-1}$ in the Local Bubble \citep{medan2019}
with luminosities from of the blue source(s) modified by their
distances from the clouds. Overlaid are the resultant values for IC\,59 and IC\,63 corrected for the Sh 2-185 geometry (Caputo et al. 2020, in press). As is evident from Figure \ref{Fig:fig1_soam17} some of the dust in Sh 2-185 extends beyond the immediate confines of the two clouds.  The exact location of this dust is not clear, but given the dynamics of H\, II regions it is likely to either be associated with the dense clouds, or swept up in an irregular shell around the star.  While this means that the exact level of the radiation field strength at the location of this dust is also unclear, we assign each of these line of sight to the nearest of the two clouds and, treat the unknown difference in radiation field strength as systematic errors.  As discussed above, systematically excluding individual data points from the analysis does not change our results, indicating that these uncertainties do not dominate the behaviour of the grain alignment in Sh 2-185. Finally, the alignment efficiency and estimated illumination flux based on the data for LDN\ 204 from \citet{cashman2014} are also plotted.


As seen in Figure \ref{Fig:comp_bubb}, for the Local Bubble wall data alone \citep{medan2019} found a best linear fit of $A = 0.067\pm0.012$ and $B = 0.0047\pm0.0002$. Including the results for IC\ 59 and IC\ 63 yields parameter values of $A = 0.087\pm0.011$ and $B = 0.0044\pm0.0002$, within the 2$\sigma$ mutual errors of the fits for the Local Bubble wall alone. If we include the results for LDN\,204- Cloud 3 \citep{cashman2014}, a significantly shallower slope results, but with a poorer fit to the low luminosity end, with parameters  $A = 0.347\pm0.023$ and $B = 0.0013\pm0.0001$.   
  
Because not all grains can cause polarization (due to sphericity, grain mineralogy, etc.), we would, under RAT alignment, expect that there will exist a radiative illumination intensity where the observed linear relationship between polarization efficiency and radiation intensity breaks down, since at that intensity all grains that \textit{can} be aligned \textit{will} be aligned. It is possible that the change in slope seen when including the LDN\ 204-Cloud 3 \citep{cashman2014} results is an indication of such a saturation effect. We are in the process of exploring the polarization efficiency in even more extreme environment where this possibility can be further tested.

\subsection{Comparison with modeled values}

\citet{Hoang2015} analytically modeled the grain alignment by radiative
torque in the IC\,63 nebula.  The presence of external high energy
radiation provides extra torque to the dust grains providing opportunity
to quantify the RAT and effects of additional torque.
\citet{Hoang2015} noticed the shallower slope, i.e.
$\alpha\,\sim\,-0.1$ for $A_{V} < 3$ and a very steep slope of
$\alpha\,\sim\,-2$ for $A_{V} > 4$.  They attributed the enhancement in
$P_{\lambda}/A_{V}$ due to combined effects of extra radiative torque
and H$_2$ formation torque.  The high grain alignment and polarization
efficiency in IC\,63 also implies the lowest values of rotational
of anisotropic radiation field compared to the values in Local Bubble.
The observed polarization values in IC\,59 and IC\,63 probe the regions
of $A_V < 3.5$, and we obtained the slopes of --0.55 and --0.41 in
IC\,63 and IC\,59, respectively.  This value is in good agreement with the modeled values of \citet{Hoang2015}.

\section{Summary}\label{sec:conc}

We used the optical polarization and multiband photometric observations
towards nebulae IC\,59 and IC\,63 to probe the grain alignment and
polarization efficiency. Using the distances to the nebulae and known
flux of their illuminating star $\gamma$ Cas we have interpreted the
results in the context of radiative grain alignment.  We compared our
results to polarization efficiencies measured in the Local Bubble walls
\citep{medan2019}.  Polarization efficiencies are found to be higher in
the IC\,59 and IC\,63 nebulae than those estimated for the Local Bubble.
This suggests an enhanced grain alignment due to RAT in these
nebulae \enquote{because of the close} vicinity of $\gamma$ Cas compared to the large distances of the walls of Local Bubble from OB associations.  Because not all grains can cause polarization (due to sphericity, grain mineralogy, etc.), we would expect that there will exist a radiative intensity where the observed
linear relationship between polarization efficiency and radiation
intensity breaks down.  We are currently exploring such extreme environments.

\bigskip
\bigskip
\bigskip
We thank anonymous referee for very constructive reports which helped significantly in improvement in the content of this paper. The WHT is operated on the island of La Palma by the Isaac Newton Group of Telescopes in the Spanish Observatorio del Roque de Los Muchachos of the Instituto de Astrofísica de Canarias. B-G.A. and A.S. acknowledge the financial support from
the NSF through grant AST-1715876. V.S., A.K. and J.Z. acknowledge
the financial support from the Research Council of Lithuania, grant No.
S-MIP-17-74.

\facility{WHT, AIMPOL, VATT, Mol\.etai}


\software{Astropy \citep{2013A&A...558A..33A}, Matplotlib \citep{Hunter:2007}, SciPy \citep{2020SciPy-NMeth}, NumPy \citep{2020NumPy-Array}}

\bibliography{bgbiblio_tot}

\begin{thebibliography}{}
\expandafter\ifx\csname natexlab\endcsname\relax\def\natexlab#1{#1}\fi
\providecommand{\url}[1]{\href{#1}{#1}}
\providecommand{\dodoi}[1]{doi:~\href{http://doi.org/#1}{\nolinkurl{#1}}}
\providecommand{\doeprint}[1]{\href{http://ascl.net/#1}{\nolinkurl{http://ascl.net/#1}}}
\providecommand{\doarXiv}[1]{\href{https://arxiv.org/abs/#1}{\nolinkurl{https://arxiv.org/abs/#1}}}

\bibitem[{{Alves} {et~al.}(2014){Alves}, {Frau}, {Girart}, {Franco}, {Santos},
  \& {Wiesemeyer}}]{alves2014}
{Alves}, F.~O., {Frau}, P., {Girart}, J.~M., {et~al.} 2014, \aap, 569, L1,
  \dodoi{10.1051/0004-6361/201424678}

\bibitem[{{Andersson} {et~al.}(2018){Andersson}, {Hoang}, {Lopez-Rodriguez},
  {Vaillancourt}, {Sankrit}, {Lazarian}, \& {HAWC+ Instrument
  Team}}]{2018AAS...23141404A}
{Andersson}, B.~G., {Hoang}, T., {Lopez-Rodriguez}, E., {et~al.} 2018, in
  American Astronomical Society Meeting Abstracts, Vol. 231, American
  Astronomical Society Meeting Abstracts \#231, 414.04

\bibitem[{{Andersson} {et~al.}(2015){Andersson}, {Lazarian}, \&
  {Vaillancourt}}]{bga2015b}
{Andersson}, B.-G., {Lazarian}, A., \& {Vaillancourt}, J.~E. 2015, \araa, 53,
  501, \dodoi{10.1146/annurev-astro-082214-122414}

\bibitem[{{Andersson} \& {Potter}(2007)}]{bga2007}
{Andersson}, B.-G., \& {Potter}, S.~B. 2007, \apj, 665, 369,
  \dodoi{10.1086/519755}

\bibitem[{{Andersson} {et~al.}(2013){Andersson}, {Piirola}, {De Buizer},
  {Clemens}, {Uomoto}, {Charcos-Llorens}, {Geballe}, {Lazarian}, {Hoang}, \&
  {Vornanen}}]{bga2013}
{Andersson}, B.-G., {Piirola}, V., {De Buizer}, J., {et~al.} 2013, \apj, 775,
  84, \dodoi{10.1088/0004-637X/775/2/84}

\bibitem[{{Astropy Collaboration} {et~al.}(2013){Astropy Collaboration},
  {Robitaille}, {Tollerud}, {Greenfield}, {Droettboom}, {Bray}, {Aldcroft},
  {Davis}, {Ginsburg}, {Price-Whelan}, {Kerzendorf}, {Conley}, {Crighton},
  {Barbary}, {Muna}, {Ferguson}, {Grollier}, {Parikh}, {Nair}, {Unther},
  {Deil}, {Woillez}, {Conseil}, {Kramer}, {Turner}, {Singer}, {Fox}, {Weaver},
  {Zabalza}, {Edwards}, {Azalee Bostroem}, {Burke}, {Casey}, {Crawford},
  {Dencheva}, {Ely}, {Jenness}, {Labrie}, {Lim}, {Pierfederici}, {Pontzen},
  {Ptak}, {Refsdal}, {Servillat}, \& {Streicher}}]{2013A&A...558A..33A}
{Astropy Collaboration}, {Robitaille}, T.~P., {Tollerud}, E.~J., {et~al.} 2013,
  \aap, 558, A33, \dodoi{10.1051/0004-6361/201322068}

\bibitem[{{Bailer-Jones} {et~al.}(2018){Bailer-Jones}, {Rybizki}, {Fouesneau},
  {Mantelet}, \& {Andrae}}]{bailer2018}
{Bailer-Jones}, C.~A.~L., {Rybizki}, J., {Fouesneau}, M., {Mantelet}, G., \&
  {Andrae}, R. 2018, \aj, 156, 58, \dodoi{10.3847/1538-3881/aacb21}

\bibitem[{{Berdyugin} {et~al.}(2014){Berdyugin}, {Piirola}, \&
  {Teerikorpi}}]{berdyugin2014}
{Berdyugin}, A., {Piirola}, V., \& {Teerikorpi}, P. 2014, \aap, 561, A24,
  \dodoi{10.1051/0004-6361/201322604}

\bibitem[{{Bhatt} \& {Jain}(1993)}]{bhatt1993}
{Bhatt}, H.~C., \& {Jain}, S.~K. 1993, \aap, 276, 507

\bibitem[{{Cashman} \& {Clemens}(2014)}]{cashman2014}
{Cashman}, L.~R., \& {Clemens}, D.~P. 2014, \apj, 793, 126,
  \dodoi{10.1088/0004-637X/793/2/126}

\bibitem[{{Chapman} {et~al.}(2011){Chapman}, {Goldsmith}, {Pineda}, {Clemens},
  {Li}, \& {Kr{\v{c}}o}}]{chapman2011}
{Chapman}, N.~L., {Goldsmith}, P.~F., {Pineda}, J.~L., {et~al.} 2011, \apj,
  741, 21, \dodoi{10.1088/0004-637X/741/1/21}

\bibitem[{{Crutcher} {et~al.}(2003){Crutcher}, {Heiles}, \&
  {Troland}}]{crutcher2003}
{Crutcher}, R., {Heiles}, C., \& {Troland}, T. 2003, in Lecture Notes in
  Physics, Berlin Springer Verlag, Vol. 614, Turbulence and Magnetic Fields in
  Astrophysics, ed. E.~{Falgarone} \& T.~{Passot}, 155--181

\bibitem[{{Cudlip} {et~al.}(1982){Cudlip}, {Furniss}, {King}, \&
  {Jennings}}]{cudlip1982}
{Cudlip}, W., {Furniss}, I., {King}, K.~J., \& {Jennings}, R.~E. 1982, \mnras,
  200, 1169, \dodoi{10.1093/mnras/200.4.1169}

\bibitem[{{Dame} {et~al.}(2001){Dame}, {Hartmann}, \& {Thaddeus}}]{dame2001}
{Dame}, T.~M., {Hartmann}, D., \& {Thaddeus}, P. 2001, \apj, 547, 792,
  \dodoi{10.1086/318388}

\bibitem[{{Davis} \& {Greenstein}(1951)}]{davis1951a}
{Davis}, L.~J., \& {Greenstein}, J.~L. 1951, \apj, 114, 206

\bibitem[{{Dolginov} \& {Mitrofanov}(1976)}]{dolginov1976}
{Dolginov}, A.~Z., \& {Mitrofanov}, I.~G. 1976, \apss, 43, 291

\bibitem[{{Dotson} {et~al.}(2000){Dotson}, {Davidson}, {Dowell}, {Schleuning},
  \& {Hildebrand}}]{dotson2000}
{Dotson}, J.~L., {Davidson}, J., {Dowell}, C.~D., {Schleuning}, D.~A., \&
  {Hildebrand}, R.~H. 2000, \apjs, 128, 335, \dodoi{10.1086/313384}

\bibitem[{{Draine} \& {Weingartner}(1996)}]{draine1996}
{Draine}, B.~T., \& {Weingartner}, J.~C. 1996, \apj, 470, 551,
  \dodoi{10.1086/177887}

\bibitem[{{Gendelev} \& {Krumholz}(2012)}]{gendelev2012}
{Gendelev}, L., \& {Krumholz}, M.~R. 2012, \apj, 745, 158,
  \dodoi{10.1088/0004-637X/745/2/158}

\bibitem[{{Hall}(1949)}]{hall1949}
{Hall}, J.~S. 1949, Science, 109, 166

\bibitem[{Harris {et~al.}(2020)Harris, Millman, van~der Walt, Gommers,
  Virtanen, Cournapeau, Wieser, Taylor, Berg, Smith, Kern, Picus, Hoyer, van
  Kerkwijk, Brett, Haldane, Fernández~del Río, Wiebe, Peterson,
  Gérard-Marchant, Sheppard, Reddy, Weckesser, Abbasi, Gohlke, \&
  Oliphant}]{2020NumPy-Array}
Harris, C.~R., Millman, K.~J., van~der Walt, S.~J., {et~al.} 2020, Nature, 585,
  357–362, \dodoi{10.1038/s41586-020-2649-2}

\bibitem[{{Heyer} {et~al.}(1998){Heyer}, {Brunt}, {Snell}, {Howe}, {Schloerb},
  \& {Carpenter}}]{heyer1998ApJS}
{Heyer}, M.~H., {Brunt}, C., {Snell}, R.~L., {et~al.} 1998, \apjs, 115, 241,
  \dodoi{10.1086/313086}

\bibitem[{{Hiltner}(1949{\natexlab{a}})}]{hiltner1949a}
{Hiltner}, W.~A. 1949{\natexlab{a}}, Science, 109, 165

\bibitem[{{Hiltner}(1949{\natexlab{b}})}]{hiltner1949b}
---. 1949{\natexlab{b}}, \apj, 109, 471

\bibitem[{{Hoang} {et~al.}(2015){Hoang}, {Lazarian}, \&
  {Andersson}}]{Hoang2015}
{Hoang}, T., {Lazarian}, A., \& {Andersson}, B.-G. 2015, \mnras, 448, 1178,
  \dodoi{10.1093/mnras/stu2758}

\bibitem[{{Hodapp}(1987)}]{hodapp1987}
{Hodapp}, K.-W. 1987, \apj, 319, 842, \dodoi{10.1086/165502}

\bibitem[{{H{\o}g} {et~al.}(2000){H{\o}g}, {Fabricius}, {Makarov}, {Urban},
  {Corbin}, {Wycoff}, {Bastian}, {Schwekendiek}, \& {Wicenec}}]{tycho2000}
{H{\o}g}, E., {Fabricius}, C., {Makarov}, V.~V., {et~al.} 2000, \aap, 355, L27.
\newblock
  \url{http://cdsads.u-strasbg.fr/cgi-bin/nph-bib_query?bibcode=2000A%26A...355L..27H&db_key=AST}

\bibitem[{{Hough} {et~al.}(2008){Hough}, {Aitken}, {Whittet}, {Adamson}, \&
  {Chrysostomou}}]{hough2008}
{Hough}, J.~H., {Aitken}, D.~K., {Whittet}, D.~C.~B., {Adamson}, A.~J., \&
  {Chrysostomou}, A. 2008, \mnras, 387, 797,
  \dodoi{10.1111/j.1365-2966.2008.13274.x}

\bibitem[{Hunter(2007)}]{Hunter:2007}
Hunter, J.~D. 2007, Computing in Science \& Engineering, 9, 90,
  \dodoi{10.1109/MCSE.2007.55}

\bibitem[{{Jansen} {et~al.}(1994){Jansen}, {van Dishoeck}, \&
  {Black}}]{jansen1994}
{Jansen}, D.~J., {van Dishoeck}, E.~F., \& {Black}, J.~H. 1994, \aap, 282, 605

\bibitem[{{Jones} {et~al.}(2015){Jones}, {Bagley}, {Krejny}, {Andersson}, \&
  {Bastien}}]{Jones2015}
{Jones}, T.~J., {Bagley}, M., {Krejny}, M., {Andersson}, B.-G., \& {Bastien},
  P. 2015, \aj, 149, 31, \dodoi{10.1088/0004-6256/149/1/31}

\bibitem[{{Jones} {et~al.}(1992){Jones}, {Klebe}, \& {Dickey}}]{jones1992}
{Jones}, T.~J., {Klebe}, D., \& {Dickey}, J.~M. 1992, \apj, 389, 602,
  \dodoi{10.1086/171233}

\bibitem[{{Karr} {et~al.}(2005){Karr}, {Noriega-Crespo}, \&
  {Martin}}]{karr2005}
{Karr}, J.~L., {Noriega-Crespo}, A., \& {Martin}, P.~G. 2005, \aj, 129, 954,
  \dodoi{10.1086/426912}

\bibitem[{{Kataoka} {et~al.}(2017){Kataoka}, {Tsukagoshi}, {Pohl}, {Muto},
  {Nagai}, {Stephens}, {Tomisaka}, \& {Momose}}]{2017ApJ...844L...5K}
{Kataoka}, A., {Tsukagoshi}, T., {Pohl}, A., {et~al.} 2017, \apjl, 844, L5,
  \dodoi{10.3847/2041-8213/aa7e33}

\bibitem[{{Lallement} {et~al.}(2014){Lallement}, {Vergely}, {Valette},
  {Puspitarini}, {Eyer}, \& {Casagrande}}]{lallement2014}
{Lallement}, R., {Vergely}, J.-L., {Valette}, B., {et~al.} 2014, \aap, 561,
  A91, \dodoi{10.1051/0004-6361/201322032}

\bibitem[{{Lallement} {et~al.}(2003){Lallement}, {Welsh}, {Vergely}, {Crifo},
  \& {Sfeir}}]{lallement2003}
{Lallement}, R., {Welsh}, B.~Y., {Vergely}, J.~L., {Crifo}, F., \& {Sfeir}, D.
  2003, \aap, 411, 447

\bibitem[{{Lazarian} \& {Draine}(1999)}]{lazarian1999b}
{Lazarian}, A., \& {Draine}, B.~T. 1999, \apjl, 516, L37,
  \dodoi{10.1086/311986}

\bibitem[{{Lazarian} \& {Hoang}(2007)}]{lazarian2007}
{Lazarian}, A., \& {Hoang}, T. 2007, \mnras, 378, 910,
  \dodoi{10.1111/j.1365-2966.2007.11817.x}

\bibitem[{{Lazarian} \& {Hoang}(2019)}]{2019ApJ...883..122L}
---. 2019, \apj, 883, 122, \dodoi{10.3847/1538-4357/ab3d39}

\bibitem[{{Matthews} {et~al.}(2009){Matthews}, {McPhee}, {Fissel}, \&
  {Curran}}]{matthews2009}
{Matthews}, B.~C., {McPhee}, C.~A., {Fissel}, L.~M., \& {Curran}, R.~L. 2009,
  \apjs, 182, 143, \dodoi{10.1088/0067-0049/182/1/143}

\bibitem[{{Medan} \& {Andersson}(2019)}]{medan2019}
{Medan}, I., \& {Andersson}, B.-G. 2019, \apj, 873, 87,
  \dodoi{10.3847/1538-4357/ab063c}

\bibitem[{{Navarete} {et~al.}(2019){Navarete}, {Galli}, \&
  {Damineli}}]{2019MNRAS.487.2771N}
{Navarete}, F., {Galli}, P. A.~B., \& {Damineli}, A. 2019, \mnras, 487, 2771,
  \dodoi{10.1093/mnras/stz1442}

\bibitem[{{Pereyra} \& {Magalh{\~a}es}(2002)}]{pereyra2002}
{Pereyra}, A., \& {Magalh{\~a}es}, A.~M. 2002, \apjs, 141, 469,
  \dodoi{10.1086/340933}

\bibitem[{{Purcell}(1979)}]{purcell1979}
{Purcell}, E.~M. 1979, \apj, 231, 404, \dodoi{10.1086/157204}

\bibitem[{{Rao} {et~al.}(1998){Rao}, {Crutcher}, {Plambeck}, \&
  {Wright}}]{rao1998}
{Rao}, R., {Crutcher}, R.~M., {Plambeck}, R.~L., \& {Wright}, M.~C.~H. 1998,
  \apjl, 502, L75, \dodoi{10.1086/311485}

\bibitem[{{Reid} {et~al.}(2009){Reid}, {Menten}, {Zheng}, {Brunthaler},
  {Moscadelli}, {Xu}, {Zhang}, {Sato}, {Honma}, {Hirota}, {Hachisuka}, {Choi},
  {Moellenbrock}, \& {Bartkiewicz}}]{reid2009}
{Reid}, M.~J., {Menten}, K.~M., {Zheng}, X.~W., {et~al.} 2009, \apj, 700, 137,
  \dodoi{10.1088/0004-637X/700/1/137}

\bibitem[{{Serkowski}(1973)}]{serkowski1973}
{Serkowski}, K. 1973, in IAU Symp. 52: Interstellar Dust and Related Topics,
  Vol.~52, 145.
\newblock
  \url{http://adsabs.harvard.edu/cgi-bin/nph-bib_query?bibcode=1973IAUS...52..145S&db_key=AST}

\bibitem[{{Soam} {et~al.}(2017){Soam}, {Maheswar}, {Lee}, {Neha}, \&
  {Andersson}}]{soam2017}
{Soam}, A., {Maheswar}, G., {Lee}, C.~W., {Neha}, S., \& {Andersson}, B.~G.
  2017, \mnras, 465, 559, \dodoi{10.1093/mnras/stw2649}

\bibitem[{{Stil} {et~al.}(2009){Stil}, {Wityk}, {Ouyed}, \&
  {Taylor}}]{stil2009}
{Stil}, J., {Wityk}, N., {Ouyed}, R., \& {Taylor}, A.~R. 2009, \apj, 701, 330,
  \dodoi{10.1088/0004-637X/701/1/330}

\bibitem[{{Strai{\v{z}}ys}(1992)}]{Straizys1992}
{Strai{\v{z}}ys}, V. 1992, {Multicolor stellar photometry (Tucson, Arizona,
  Pachart Publishing House), available in pdf format from
  http://www.itpa.lt/MulticolorStellarPhotometry/}

\bibitem[{{Strai{\v{z}}ys} {et~al.}(2013){Strai{\v{z}}ys}, {Boyle}, {Janusz},
  {Laugalys}, \& {Kazlauskas}}]{Straizys2013}
{Strai{\v{z}}ys}, V., {Boyle}, R.~P., {Janusz}, R., {Laugalys}, V., \&
  {Kazlauskas}, A. 2013, \aap, 554, A3, \dodoi{10.1051/0004-6361/201321029}

\bibitem[{{Strai{\v{z}}ys} {et~al.}(2019){Strai{\v{z}}ys}, {Boyle},
  {Mila{\v{s}}ius}, {{\v{C}}ernis}, {Macijauskas}, {Munari}, {Janusz},
  {Zdanavi{\v{c}}ius}, {Zdanavi{\v{c}}ius}, {Maskoli{\={u}}nas},
  {Raudeli{\={u}}nas}, \& {Kazlauskas}}]{Straizys2019}
{Strai{\v{z}}ys}, V., {Boyle}, R.~P., {Mila{\v{s}}ius}, K., {et~al.} 2019,
  \aap, 623, A22, \dodoi{10.1051/0004-6361/201833987}

\bibitem[{{Sugitani} {et~al.}(2011){Sugitani}, {Nakamura}, {Watanabe},
  {Tamura}, {Nishiyama}, {Nagayama}, {Kandori}, {Nagata}, {Sato}, {Gutermuth},
  {Wilson}, \& {Kawabe}}]{sugitani2011}
{Sugitani}, K., {Nakamura}, F., {Watanabe}, M., {et~al.} 2011, \apj, 734, 63,
  \dodoi{10.1088/0004-637X/734/1/63}

\bibitem[{{Vaillancourt} {et~al.}(2020){Vaillancourt}, {Andersson}, {Clemens},
  {Piirola}, {Hoang}, {Becklin}, \& {Caputo}}]{vaillancourt2020}
{Vaillancourt}, J.~E., {Andersson}, B.-G., {Clemens}, D.~P., {et~al.} 2020,
  arXiv e-prints, arXiv:2011.00114.
\newblock \doarXiv{2011.00114}

\bibitem[{{Vaillancourt} \& {Matthews}(2012)}]{vaillancourt2012}
{Vaillancourt}, J.~E., \& {Matthews}, B.~C. 2012, \apjs, 201, 13,
  \dodoi{10.1088/0067-0049/201/2/13}

\bibitem[{Virtanen {et~al.}(2020)Virtanen, Gommers, Oliphant, Haberland, Reddy,
  Cournapeau, Burovski, Peterson, Weckesser, Bright, {van der Walt}, Brett,
  Wilson, Millman, Mayorov, Nelson, Jones, Kern, Larson, Carey, Polat, Feng,
  Moore, {VanderPlas}, Laxalde, Perktold, Cimrman, Henriksen, Quintero, Harris,
  Archibald, Ribeiro, Pedregosa, {van Mulbregt}, \& {SciPy 1.0
  Contributors}}]{2020SciPy-NMeth}
Virtanen, P., Gommers, R., Oliphant, T.~E., {et~al.} 2020, Nature Methods, 17,
  261, \dodoi{10.1038/s41592-019-0686-2}

\bibitem[{{Vrba} {et~al.}(1976){Vrba}, {Strom}, \& {Strom}}]{vrba1976}
{Vrba}, F.~J., {Strom}, S.~E., \& {Strom}, K.~M. 1976, \aj, 81, 958,
  \dodoi{10.1086/111976}

\bibitem[{{Whittet}(2003)}]{whittet2003}
{Whittet}, D. C.~B. 2003, Dust in the galactic environment - 2:nd ed. (Dust in
  the galactic environment Institute of Physics Publishing, 390 p.).
\newblock
  \url{http://adsabs.harvard.edu/cgi-bin/nph-bib_query?bibcode=2003QB791.W45......&db_key=AST}

\bibitem[{{Whittet} {et~al.}(2001){Whittet}, {Gerakines}, {Hough}, \&
  {Shenoy}}]{whittet2001}
{Whittet}, D.~C.~B., {Gerakines}, P.~A., {Hough}, J.~H., \& {Shenoy}, S.~S.
  2001, \apj, 547, 872, \dodoi{10.1086/318421}

\bibitem[{{Wright} {et~al.}(2003){Wright}, {Egan}, {Kraemer}, \&
  {Price}}]{wright2003}
{Wright}, C.~O., {Egan}, M.~P., {Kraemer}, K.~E., \& {Price}, S.~D. 2003, \aj,
  125, 359, \dodoi{10.1086/345511}

\end{thebibliography}
\bibliographystyle{aasjournal}

\startlongtable
\begin{table}
  \centering
    \caption{Results of photometry and classification of stars in the
Vilnius system for the IC\,59 stars with the VATT telescope.}\label{tab:phot59}
\resizebox{\textwidth}{!}{
\begin{tabular}{lllllllllllllllllllr}
\hline
  No.& RA (J2000)  & DEC (J2000)  & $V$& $U-V$& $P-V$& $X-V$& $Y-V$& $Z-V$& $V-S$&
$e(V)$& $e(U-V)$& $e(P-V)$& $e(X-V)$& $e(Y-V)$& $e(Z-V)$,& $e(V-S)$ & Sp type & $A_{V}$& d\\
  & (hh:mm:ss)  & (dd:mm:ss)   & (mag)&(mag)&(mag)&(mag)&(mag)&(mag)&(mag)&(mag)&(mag)&(mag)&(mag)&(mag)&(mag)&(mag)& & (mag)&(pc)\\
  \hline
   1& 00:56:37.79& +61:07:49.5& 16.341&   --  &  2.867&  1.963&  0.896&  0.313&  0.859&  0.012&   --  &  0.033&  0.017&  0.013&  0.016&  0.013& f5 III & 1.868 &7008.0     \\
   2& 00:56:38.76& +61:12:48.3& 15.831&  3.636&  2.694&  1.748&  0.788&  0.290&  0.674&  0.032&  0.062&  0.026&  0.033&  0.012&  0.039&  0.012& a6 IV  & 2.205 &2170.0     \\
   3& 00:56:43.71& +61:03:28.5& 13.669&  2.777&  2.071&  1.296&  0.544&  0.198&  0.539&  0.013&  0.020&  0.025&  0.016&  0.014&  0.017&  0.012& a5 V   & 1.24  &1102.0     \\
   4& 00:56:46.94& +61:10:13.2& 13.316&  2.974&  2.245&  1.543&  0.693&  0.270&  0.635&  0.012&  0.020&  0.025&  0.015&  0.013&  0.016&  0.012& f5 III & 1.023 &1056.0     \\
   5& 00:56:47.02& +61:07:52.1& 14.418&  3.063&  2.268&  1.535&  0.757&  0.260&  0.637&  0.013&  0.021&  0.026&  0.016&  0.016&  0.018&  0.013& f3 V   & 1.439 &2988.0     \\
   6& 00:56:47.65& +61:12:25.0& 14.963&  3.237&  2.407&  1.661&  0.831&  0.292&  0.685&  0.014&  0.022&  0.026&  0.017&  0.019&  0.021&  0.014& f3 V   & 1.747 &3160.0     \\
\end{tabular}%
}
\end{table}

 \startlongtable
\begin{table}
  \centering
    \caption{Results of photometry and classification of stars in the
Vilnius system for the IC\,63 stars with the VATT telescope.}\label{tab:phot63}
\resizebox{\textwidth}{!}{\begin{tabular}{lllllllllllllllllllr}
  No.& RA (J2000)  & DEC (J2000)  & $V$& $U-V$& $P-V$& $X-V$& $Y-V$& $Z-V$& $V-S$&
$e(V)$& $e(U-V)$& $e(P-V)$& $e(X-V)$& $e(Y-V)$& $e(Z-V)$,& $e(V-S)$ & Sp type & $A_{V}$& d\\
  & (hh:mm:ss)  & (dd:mm:ss)   & (mag)&(mag)&(mag)&(mag)&(mag)&(mag)&(mag)&(mag)&(mag)&(mag)&(mag)&(mag)&(mag)&(mag)& &(mag) & (pc) \\
  \hline
   1 &00:58:39.02& +60:55:56.4& 17.717&  3.672&   --  &  2.093&  0.979&  0.375&  0.884&  0.014&  0.063&   --  &  0.020&  0.015&  0.018&  0.016& g1 IV  &1.801 &3705.0 \\ 
   2 &00:58:40.22& +60:52:19.8& 14.451&  3.211&  2.390&  1.438&  0.649&  0.233&  0.572&  0.013&  0.022&  0.025&  0.016&  0.016&  0.018&  0.012& a2 V   &1.847 &1265.0 \\ 
   3 &00:58:40.73& +60:58:59.0& 15.417&  3.402&  2.561&  1.675&  0.766&  0.284&  0.675&  0.016&  0.027&  0.025&  0.019&  0.024&  0.026&  0.016& a7 IV  &2.051 &2416.0 \\ 
   4 &00:58:40.94& +60:56:56.2& 14.642&  6.149&  5.142&  3.623&  1.515&  0.585&  1.357&  0.013&  0.030&  0.033&  0.021&  0.022&  0.021&  0.013& k3 III &2.508 &2606.0 \\ 
   5 &00:58:41.21& +60:57:08.2& 14.825&  2.795&  2.190&  1.551&  0.667&  0.260&  0.631&  0.014&  0.022&  0.025&  0.017&  0.018&  0.020&  0.014& f8 V   &0.678 &948.0  \\ 
   6 &00:58:41.25& +60:51:00.1& 15.884&  4.997&   --  &  2.996&  1.236&  0.472&  1.187&  0.012&  0.050&   --  &  0.016&  0.012&  0.015&  0.012& k2 III &1.568 &5994.0 \\
\end{tabular}}
\end{table} 
\startlongtable
\begin{table}
\centering
\resizebox{\textwidth}{!}{\begin{tabular}{lllllllllllllllllllr}
  No.& RA (J2000)  & DEC (J2000)  & $V$& $U-V$& $P-V$& $X-V$& $Y-V$& $Z-V$& $V-S$&
$e(V)$& $e(U-V)$& $e(P-V)$& $e(X-V)$& $e(Y-V)$& $e(Z-V)$,& $e(V-S)$ & Sp type&$A_{V}$& d\\
  & (hh:mm:ss)  & (dd:mm:ss)   & (mag)&(mag)&(mag)&(mag)&(mag)&(mag)&(mag)&(mag)&(mag)&(mag)&(mag)&(mag)&(mag)&(mag)& &(mag) & (pc) \\
  \hline
  66& 00:59:16.98& +60:57:39.3& 17.100&  3.599&   --  &  2.082&  0.977&  0.378&  0.811&  0.013&  0.028&   --  &  0.016&  0.013&  0.016&  0.013& g0 V   &1.86  &1838.0 \\ 
  67& 00:59:17.61& +60:52:02.6& 15.254&  3.231&  2.664&  1.812&  0.774&  0.324&  0.700&  0.015&  0.024&  0.025&  0.018&  0.022&  0.024&  0.015& g5 V   &0.786 &1027.0 \\ 
  68& 00:59:17.73& +60:51:26.9& 17.194&  3.245&   --  &  1.796&  0.827&  0.324&  0.766&  0.013&  0.025&   --  &  0.016&  0.013&  0.016&  0.013& f9 IV  &1.277 &4350.0 \\ 
  69& 00:59:18.12& +60:58:30.6& 12.585&  2.620&  1.978&  1.252&  0.532&  0.206&  0.480&  0.012&  0.020&  0.025&  0.015&  0.012&  0.015&  0.012& f2 V   &0.62  &615.0  \\ 
  70& 00:59:18.22& +60:50:52.5& 14.340&  4.260&  3.476&  2.485&  1.037&  0.411&  0.964&  0.013&  0.024&  0.026&  0.016&  0.016&  0.018&  0.013& g7 III &1.456 &2932.0 \\ 
  71& 00:59:18.52& +60:55:34.4& 13.966&  3.363&  2.771&  1.962&  0.836&  0.366&  0.786&  0.013&  0.021&  0.025&  0.016&  0.014&  0.016&  0.013& g7 V   &0.973 &291.0  \\
\end{tabular}}
\end{table} 
\startlongtable
\begin{table}
\centering
\resizebox{\textwidth}{!}{\begin{tabular}{lllllllllllllllllllr}
  No.& RA (J2000)  & DEC (J2000)  & $V$& $U-V$& $P-V$& $X-V$& $Y-V$& $Z-V$& $V-S$&
$e(V)$& $e(U-V)$& $e(P-V)$& $e(X-V)$& $e(Y-V)$& $e(Z-V)$,& $e(V-S)$ & Sp type &$A_{V}$& d\\
  & (hh:mm:ss)  & (dd:mm:ss)   & (mag)&(mag)&(mag)&(mag)&(mag)&(mag)&(mag)&(mag)&(mag)&(mag)&(mag)&(mag)&(mag)&(mag)& &(mag) & (pc) \\
  \hline
 131& 00:59:51.38& +60:50:09.8& 17.486&  3.464&   --  &  2.006&  0.906&  0.325&  0.768&  0.014&  0.042&   --  &  0.018&  0.014&  0.017&  0.014& g1 IV  &1.498 &2767.0 \\ 
 132& 00:59:51.86& +60:50:21.7& 17.584&  3.256&   --  &  1.920&  0.864&  0.312&  0.800&  0.014&  0.033&   --  &  0.018&  0.014&  0.017&  0.015& g3 IV  &1.219 &4094.0 \\ 
 133& 00:59:52.05& +60:57:46.5& 13.680&  4.115&  3.464&  2.375&  0.961&  0.389&  0.908&  0.012&  0.020&  0.025&  0.015&  0.012&  0.015&  0.012& k0 III &0.79  &1123.0 \\ 
 134& 00:59:52.23& +60:51:30.6& 14.544&  2.584&  1.840&  1.024&  0.484&  0.185&  0.427&  0.012&  0.020&  0.025&  0.015&  0.012&  0.015&  0.012& a1 V   &1.215 &2722.0 \\ 
 135& 00:59:52.44& +60:57:27.9& 14.448&  2.452&  1.924&  1.329&  0.646&  0.237&  0.559&  0.012&  0.020&  0.025&  0.015&  0.012&  0.015&  0.012& f6 V   &0.72  &3259.0 \\ 
 136& 00:59:52.66& +60:49:32.8& 16.782&   --  &  2.288&  1.550&  0.701&  0.252&  0.627&  0.012&   --  &  0.030&  0.016&  0.013&  0.016&  0.013& f5 V   &1.023 &2897.0 \\
\hline                                                                                                                               
\end{tabular}}                                                                                                    
\end{table}

\startlongtable
\begin{table}
  \centering
    \caption{ Results of photometry and classification of stars in the
Vilnius system for the IC\,59 and IC\,63 stars with the Maksutov-type telescope}\label{tab:phot5963}
\resizebox{\textwidth}{!}{\begin{tabular}{lllllllllllllllllllr}
\hline
  No.& RA (J2000)  & DEC (J2000)  & $V$& $U-V$& $P-V$& $X-V$& $Y-V$& $Z-V$& $V-S$&
$e(V)$& $e(U-V)$& $e(P-V)$& $e(X-V)$& $e(Y-V)$& $e(Z-V)$,& $e(V-S)$ & Sp type & $A_{V}$& d\\
  & (hh:mm:ss)  & (dd:mm:ss)   & (mag)&(mag)&(mag)&(mag)&(mag)&(mag)&(mag)&(mag)&(mag)&(mag)&(mag)&(mag)&(mag)&(mag)& &(mag)&(pc) \\
  \hline
   1& 00:56:15.50& +61:20:35.5& 14.864&   --  &  2.899&  2.103&  1.063&  0.407&  0.849&  0.071&   --  &  0.118&  0.100&  0.113&  0.110&  0.118& f5 III& 2.563 &1405.0   \\ 
   2& 00:56:19.17& +61:06:30.2& 13.715&  2.885&  2.292&  1.604&  0.686&  0.291&  0.630&  0.028&  0.060&  0.052&  0.044&  0.035&  0.034&  0.036& g0 IV & 0.636 &908.0    \\ 
   3& 00:56:19.37& +61:16:50.7& 12.919&  2.701&  2.077&  1.396&  0.605&  0.249&  0.591&  0.026&  0.048&  0.041&  0.033&  0.031&  0.029&  0.031& f6 IV & 0.545 &767.0    \\ 
   4& 00:56:20.86& +61:11:13.6& 14.953&   --  &  2.503&  1.751&  0.769&  0.315&  0.700&  0.035&   --  &  0.079&  0.052&  0.044&  0.048&  0.045& g5 V  & 0.765 &811.0    \\ 
   5& 00:56:22.21& +61:04:37.2& 14.292&   --  &  3.096&  1.973&  0.953&  0.348&  0.875&  0.033&   --  &  0.065&  0.047&  0.042&  0.042&  0.040& a3 III& 3.087 &2434.0   \\ 
   6& 00:56:22.45& +61:20:29.4& 11.068&   --  &   --  &  1.867&  0.910&  0.379&  0.852&  0.068&   --  &   --  &  0.086&  0.088&  0.082&  0.091& a1 III& 0.562 &1048.0   \\
\end{tabular}}
\end{table} 
\startlongtable
\begin{table}
\centering
\resizebox{\textwidth}{!}{\begin{tabular}{lllllllllllllllllllr}
\hline
  No.& RA (J2000)  & DEC (J2000)  & $V$& $U-V$& $P-V$& $X-V$& $Y-V$& $Z-V$& $V-S$&
$e(V)$& $e(U-V)$& $e(P-V)$& $e(X-V)$& $e(Y-V)$& $e(Z-V)$,& $e(V-S)$ & Sp type & $A_{V}$& d\\
  & (hh:mm:ss)  & (dd:mm:ss)   & (mag)&(mag)&(mag)&(mag)&(mag)&(mag)&(mag)&(mag)&(mag)&(mag)&(mag)&(mag)&(mag)&(mag)& &(mag)&(pc) \\
  \hline
  66& 00:57:21.68& +60:59:20.1& 14.696&   --  &  2.408&  1.402&  0.658&  0.239&  0.552&  0.034&   --  &  0.064&  0.047&  0.043&  0.045&  0.044& a3 III  &1.86  &3198.0   \\ 
  67& 00:57:22.41& +61:02:11.8& 14.905&   --  &  2.271&  1.620&  0.680&  0.275&  0.615&  0.035&   --  &  0.070&  0.050&  0.043&  0.048&  0.047& g1 V    &0.574 &878.0    \\ 
  68& 00:57:23.45& +60:50:40.2& 15.337&   --  &  2.521&  1.781&  0.693&  0.286&  0.704&  0.041&   --  &  0.087&  0.058&  0.054&  0.057&  0.057& g9 V    &0.287 &875.0    \\ 
  69& 00:57:23.81& +61:22:46.8& 13.928&  3.427&  2.448&  1.490&  0.689&  0.254&  0.635&  0.029&  0.081&  0.056&  0.043&  0.037&  0.038&  0.036& a8 III  &1.656 &1957.0   \\ 
  70& 00:57:24.98& +61:01:08.6& 14.042&  2.773&  2.272&  1.548&  0.659&  0.266&  0.625&  0.030&  0.068&  0.051&  0.039&  0.037&  0.038&  0.038& g3 V    &0.391 &627.0    \\ 
  71& 00:57:26.45& +60:58:07.5& 14.746&  2.879&  2.468&  1.706&  0.743&  0.301&  0.646&  0.033&  0.098&  0.064&  0.048&  0.045&  0.046&  0.044& g4 V    &0.695 &933.0    \\
\end{tabular}}
\end{table} 
\startlongtable
\begin{table}
\centering
\resizebox{\textwidth}{!}{\begin{tabular}{lllllllllllllllllllr}
\hline
  No.& RA (J2000)  & DEC (J2000)  & $V$& $U-V$& $P-V$& $X-V$& $Y-V$& $Z-V$& $V-S$&
$e(V)$& $e(U-V)$& $e(P-V)$& $e(X-V)$& $e(Y-V)$& $e(Z-V)$,& $e(V-S)$ & Sp type & $A_{V}$& d\\
  & (hh:mm:ss)  & (dd:mm:ss)   & (mag)&(mag)&(mag)&(mag)&(mag)&(mag)&(mag)&(mag)&(mag)&(mag)&(mag)&(mag)&(mag)&(mag)&&(mag)&(pc) \\
  \hline
 131& 00:58:19.57& +60:48:13.4& 15.803&   --  &  2.010&  1.134&  0.626&  0.204&  0.505&  0.049&   --  &  0.112&  0.070&  0.066&  0.070&  0.064& b9 V   &1.947 &3224.0   \\
 132& 00:58:20.28& +60:48:47.9& 15.819&   --  &  2.389&  1.653&  0.738&  0.269&  0.661&  0.049&   --  &  0.103&  0.073&  0.096&  0.074&  0.064& g0 IV  &0.853 &2279.0   \\
 133& 00:58:20.78& +61:00:12.7& 14.177&  2.722&  2.098&  1.479&  0.654&  0.251&  0.575&  0.031&  0.071&  0.053&  0.041&  0.038&  0.040&  0.039& f5 V   &0.828 &778.0    \\
 134& 00:58:21.13& +60:52:14.8& 14.923&   --  &  2.782&  1.867&  0.820&  0.362&  0.735&  0.037&   --  &  0.075&  0.054&  0.052&  0.052&  0.047& g8 V   &0.869 &656.0    \\
 135& 00:58:21.25& +60:47:15.5& 13.619&   --  &   --  &  2.844&  1.004&  0.601&  1.065&  0.040&   --  &   --  &  0.091&  0.055&  0.061&  0.050& f6 V   &0.774 &738.0    \\
 136& 00:58:21.62& +61:09:39.1& 13.311&  2.755&  2.160&  1.483&  0.637&  0.264&  0.600&  0.026&  0.058&  0.041&  0.035&  0.033&  0.031&  0.032& f7 V   &0.616 &612.0    \\
\end{tabular}}
\end{table} 
\startlongtable
\begin{table}
\centering
\resizebox{\textwidth}{!}{\begin{tabular}{lllllllllllllllllllr}
\hline
  No.& RA (J2000)  & DEC (J2000)  & $V$& $U-V$& $P-V$& $X-V$& $Y-V$& $Z-V$& $V-S$&
$e(V)$& $e(U-V)$& $e(P-V)$& $e(X-V)$& $e(Y-V)$& $e(Z-V)$,& $e(V-S)$ & Sp type & $A_{V}$& d\\
  & (hh:mm:ss)  & (dd:mm:ss)   & (mag)&(mag)&(mag)&(mag)&(mag)&(mag)&(mag)&(mag)&(mag)&(mag)&(mag)&(mag)&(mag)&(mag)&&(mag)&(pc) \\
  \hline
 196& 00:58:57.80& +61:07:43.3& 11.447&  2.370&  1.627&  0.818&  0.339&  0.163&  0.283&  0.024&  0.038&  0.033&  0.030&  0.028&  0.026&  0.029& a3 V    & 0.499& 781.0   \\
 197& 00:58:58.13& +61:08:28.4& 13.625&  3.150&  2.430&  1.641&  0.721&  0.279&  0.663&  0.028&  0.067&  0.048&  0.037&  0.035&  0.033&  0.034& f0 IV   & 1.639& 977.0   \\
 198& 00:58:58.28& +61:09:43.3& 10.643&  2.738&  1.991&  1.177&  0.485&  0.212&  0.436&  0.023&  0.038&  0.033&  0.030&  0.028&  0.025&  0.029& a6 III  & 0.948& 470.0   \\
 199& 00:58:58.36& +61:23:42.5& 14.757&   --  &  2.641&  1.609&  0.803&  0.308&  0.645&  0.033&   --  &  0.065&  0.047&  0.043&  0.046&  0.047& a3 III  & 2.463& 4007.0  \\
 200& 00:58:58.64& +61:19:36.9& 14.146&  2.880&  2.311&  1.623&  0.693&  0.275&  0.661&  0.031&  0.092&  0.053&  0.041&  0.042&  0.039&  0.040& g3 V    & 0.532& 681.0   \\
 201& 00:58:58.72& +61:08:37.8& 15.134&   --  &  2.390&  1.646&  0.726&  0.275&  0.687&  0.047&   --  &  0.075&  0.059&  0.056&  0.060&  0.057& g1 V    & 0.765& 997.0   \\
\end{tabular}}
\end{table} 
\startlongtable
\begin{table}
\centering
\resizebox{\textwidth}{!}{\begin{tabular}{lllllllllllllllllllr}
\hline
  No.& RA (J2000)  & DEC (J2000)  & $V$& $U-V$& $P-V$& $X-V$& $Y-V$& $Z-V$& $V-S$&
$e(V)$& $e(U-V)$& $e(P-V)$& $e(X-V)$& $e(Y-V)$& $e(Z-V)$,& $e(V-S)$ & Sp type & $A_{V}$& d\\
  & (hh:mm:ss)  & (dd:mm:ss)   & (mag)&(mag)&(mag)&(mag)&(mag)&(mag)&(mag)&(mag)&(mag)&(mag)&(mag)&(mag)&(mag)&(mag)&&(mag)&(pc) \\
  \hline
 261& 00:59:43.05& +61:19:28.1& 14.278&  2.938&  2.346&  1.598&  0.703&  0.273&  0.673&  0.031&  0.099&  0.058&  0.041&  0.041&  0.041&  0.042& f9 IV  & 0.761& 1024.0   \\
 262& 00:59:43.39& +61:21:35.3& 14.142&  2.720&  2.191&  1.547&  0.664&  0.285&  0.639&  0.030&  0.074&  0.051&  0.040&  0.039&  0.038&  0.039& g1 V   & 0.508& 521.0    \\
 263& 00:59:44.30& +61:15:52.6& 12.084&  4.484&  3.757&  2.577&  1.015&  0.407&  0.945&  0.025&  0.055&  0.047&  0.033&  0.030&  0.028&  0.030& k0 III & 1.015& 1134.0   \\
 264& 00:59:44.71& +60:47:39.2& 14.988&  2.797&   --  &  1.621&  0.709&  0.275&  0.694&  0.037&  0.098&   --  &  0.054&  0.049&  0.052&  0.048& g2 V   & 0.649& 802.0    \\
 265& 00:59:45.19& +61:13:23.9& 14.728&   --  &  2.451&  1.555&  0.712&  0.295&  0.612&  0.034&   --  &  0.081&  0.048&  0.044&  0.046&  0.043& a6 V   & 1.884& 1126.0   \\
 266& 00:59:45.23& +60:48:00.7& 15.105&   --  &  2.304&  1.277&  0.562&  0.209&  0.528&  0.040&   --  &  0.076&  0.053&  0.049&  0.051&  0.051& a4 III & 1.398& 3921.0   \\
\end{tabular}}
\end{table} 
\startlongtable
\begin{table}
\centering
\resizebox{\textwidth}{!}{\begin{tabular}{lllllllllllllllllllr}
\hline
  No.& RA (J2000)  & DEC (J2000)  & $V$& $U-V$& $P-V$& $X-V$& $Y-V$& $Z-V$& $V-S$&
$e(V)$& $e(U-V)$& $e(P-V)$& $e(X-V)$& $e(Y-V)$& $e(Z-V)$,& $e(V-S)$ & Sp type & $A_{V}$& d\\
  & (hh:mm:ss)  & (dd:mm:ss)   & (mag)&(mag)&(mag)&(mag)&(mag)&(mag)&(mag)&(mag)&(mag)&(mag)&(mag)&(mag)&(mag)&(mag)&&(mag)&(pc) \\
  \hline
 326& 01:00:25.42& +60:49:10.4& 15.302&   --  &  2.349&  1.598&  0.683&  0.261&  0.714&  0.040&   --  &  0.082&  0.058&  0.051&  0.055&  0.052& g3 V  & 0.491 &1115.0   \\ 
 327& 01:00:25.64& +61:04:16.8& 15.535&   --  &  2.517&  1.698&  0.728&  0.294&  0.701&  0.045&   --  &  0.091&  0.066&  0.059&  0.060&  0.059& g5 V  & 0.595 &1030.0   \\ 
 328& 01:00:25.90& +61:21:38.4& 14.516&   --  &   --  &  1.596&  0.688&  0.251&  0.625&  0.047&   --  &   --  &  0.069&  0.063&  0.069&  0.063& k1 V  & 0.524 &427.0    \\ 
 329& 01:00:26.62& +60:51:11.2& 15.521&   --  &  2.194&  1.240&  0.577&  0.222&  0.479&  0.046&   --  &  0.100&  0.063&  0.057&  0.062&  0.063& a4 III& 1.46  &4185.0   \\ 
 330& 01:00:26.80& +60:55:08.5& 15.132&   --  &  2.253&  1.357&  0.589&  0.237&  0.563&  0.038&   --  &  0.075&  0.053&  0.049&  0.052&  0.051& a6 V  & 1.373 &2047.0   \\ 
 331& 01:00:27.77& +60:50:49.3& 15.357&   --  &  2.149&  1.185&  0.557&  0.229&  0.427&  0.041&   --  &  0.088&  0.059&  0.053&  0.054&  0.057& a4 III& 1.377 &6647.0   \\
\end{tabular}}
\end{table} 
\startlongtable
\begin{table}
\centering
\resizebox{\textwidth}{!}{\begin{tabular}{lllllllllllllllllllr}
\hline
  No.& RA (J2000)  & DEC (J2000)  & $V$& $U-V$& $P-V$& $X-V$& $Y-V$& $Z-V$& $V-S$&
$e(V)$& $e(U-V)$& $e(P-V)$& $e(X-V)$& $e(Y-V)$& $e(Z-V)$,& $e(V-S)$ & Sp type & $A_{V}$& d\\
  & (hh:mm:ss)  & (dd:mm:ss)   & (mag)&(mag)&(mag)&(mag)&(mag)&(mag)&(mag)&(mag)&(mag)&(mag)&(mag)&(mag)&(mag)&(mag)&&(mag)&(pc) \\
  \hline
 391& 01:00:51.65& +60:52:47.7& 13.830&  2.682&  2.019&  1.352&  0.578&  0.221&  0.568&  0.030&  0.064&  0.044&  0.040&  0.037&  0.038&  0.039& f2 V   & 0.811 &764.0   \\
 392& 01:00:51.95& +60:50:11.9& 13.638&  3.023&  2.080&  1.151&  0.517&  0.201&  0.469&  0.031&  0.071&  0.050&  0.038&  0.038&  0.037&  0.041& a1 V   & 1.352 &1518.0  \\
 393& 01:00:52.07& +61:09:59.2& 15.734&   --  &  2.567&  1.455&  0.677&  0.239&  0.619&  0.047&   --  &  0.106&  0.068&  0.060&  0.067&  0.065& a3 III & 1.939 &3376.0  \\
 394& 01:00:53.02& +60:50:56.0& 15.316&   --  &  2.205&  1.393&  0.610&  0.228&  0.600&  0.043&   --  &  0.084&  0.058&  0.054&  0.055&  0.057& a9 V   & 1.252 &1676.0  \\
 395& 01:00:54.20& +61:24:38.8& 13.986&   --  &  3.334&  2.160&  0.769&  0.463&  0.796&  0.030&   --  &  0.065&  0.045&  0.038&  0.035&  0.063& k4 V   & 0.029 &268.0   \\
 396& 01:00:54.63& +60:54:10.1& 14.889&   --  &  3.644&  2.464&  1.034&  0.390&  0.982&  0.038&   --  &  0.103&  0.057&  0.050&  0.051&  0.047& k0 III & 1.094 &3125.0  \\
\end{tabular}}
\end{table} 
\startlongtable
\begin{table}
\centering
\resizebox{\textwidth}{!}{\begin{tabular}{lllllllllllllllllllr}
\hline
  No.& RA (J2000)  & DEC (J2000)  & $V$& $U-V$& $P-V$& $X-V$& $Y-V$& $Z-V$& $V-S$&
$e(V)$& $e(U-V)$& $e(P-V)$& $e(X-V)$& $e(Y-V)$& $e(Z-V)$,& $e(V-S)$ & Sp type & $A_{V}$& d\\
  & (hh:mm:ss)  & (dd:mm:ss)   & (mag)&(mag)&(mag)&(mag)&(mag)&(mag)&(mag)&(mag)&(mag)&(mag)&(mag)&(mag)&(mag)&(mag)&&(mag)&(pc) \\
  \hline
 456& 01:01:15.74& +61:01:05.2& 13.928&  3.290&  2.346&  1.365&  0.646&  0.245&  0.570&  0.028&  0.084&  0.057&  0.042&  0.033&  0.036&  0.034& a2 V   &1.835 &1641.0  \\ 
 457& 01:01:15.83& +60:53:57.6& 13.991&  2.611&  1.965&  1.226&  0.533&  0.197&  0.527&  0.030&  0.070&  0.049&  0.039&  0.038&  0.038&  0.039& a8 V   &1.007 &1046.0  \\ 
 458& 01:01:16.33& +61:10:28.2& 14.920&   --  &  2.080&  1.218&  0.592&  0.242&  0.498&  0.037&   --  &  0.068&  0.049&  0.046&  0.048&  0.051& a0 V   &1.731 &2991.0  \\ 
 459& 01:01:16.68& +60:56:28.0& 15.601&   --  &   --  &  1.414&  0.566&  0.160&  0.665&  0.072&   --  &   --  &  0.099&  0.096&  0.104&  0.089& g8 V   &0.37  &1026.0  \\ 
 460& 01:01:16.83& +60:48:01.5& 16.451&   --  &  2.155&  1.228&  0.503&  0.176&  0.450&  0.065&   --  &  0.128&  0.086&  0.083&  0.114&  0.083& a2 V   &1.24  &3619.0  \\ 
 461& 01:01:16.88& +61:24:16.3& 15.725&   --  &   --  &  1.333&  0.635&  0.211&  0.546&  0.046&   --  &   --  &  0.072&  0.062&  0.064&  0.068& a0 V   &2.092 &2865.0  \\
\hline                                                                                                                                 
\end{tabular}}                                                                                                                           
\end{table}                                                                                                                             

\startlongtable
\begin{table}
  \scriptsize
  \centering
    \caption{Polarization and extinction measurements towards IC\,59 and IC\,63}\label{tab:pol_res}
\begin{tabular}{llllllllr}\hline
  No.& RA (J2000)  & DEC (J2000)  & $P\pm\sigma_{P}$ & $\theta_{pos}\pm$ $\sigma_{\theta_{pos}}$&$V_{mag}$& $A_{V}$& $d$ & $P/A_{V}$   \\
  & (hh:mm:ss)  & (dd:mm:ss)   & (\%)  & ($^{\circ}$)  & (mag) & (mag) & (pc) & (\%/mag) \\
  \hline
\multicolumn{9}{c}{\bf{IC59}}\\
   1&  00:56:38.76&   +61:12:48.3&  1.848$\pm$   0.526&   94.686$\pm$    7.840&    15.831&   2.205&   2170&    0.838   \\   
   2&  00:56:46.94&   +61:10:13.2&  2.544$\pm$   0.232&  100.928$\pm$    2.535&    13.316&   1.023&   1056&    2.487   \\   
   3&  00:56:47.02&   +61:07:52.1&  3.056$\pm$   0.398&   65.293$\pm$    3.649&    14.418&   1.439&   2988&    2.124    \\   
   4&  00:56:47.65&   +61:12:25.0&  1.148$\pm$   0.159&   49.301$\pm$    3.723&    14.963&   1.747&   3160&    0.657   \\   
   5&  00:56:54.04&   +61:09:04.2&  5.722$\pm$   1.151&   90.927$\pm$    5.863&    17.671&   3.041&   3203&    1.882    \\
\end{tabular}
\end{table} 
\startlongtable
\begin{table}
  \scriptsize
  \centering
      \begin{tabular}{llllllllr}\hline
          No.& RA (J2000)  & DEC (J2000)  & $P\pm\sigma_{P}$ & $\theta_{pos}\pm$ $\sigma_{\theta_{pos}}$&$V_{mag}$& $A_{V}$& $d$ & $P/A_{V}$   \\
  & (hh:mm:ss)  & (dd:mm:ss)   & (\%)  & ($^{\circ}$)  & (mag) & (mag) & (pc) & (\%/mag) \\
  \hline
  51&  00:57:57.43&   +61:07:36.7&  1.685$\pm$   0.651&   91.162$\pm$   10.653&    15.876&   0.549&    913&    3.069   \\   
  52&  00:57:58.76&   +61:11:14.8&  3.133$\pm$   0.091&   85.642$\pm$    0.857&    12.325&   1.756&   2331&    1.784   \\   
  53&  00:58:01.12&   +61:12:20.0&  2.278$\pm$   0.550&   86.667$\pm$    7.094&    15.704&   1.772&   3848&    1.286   \\   
  54&  00:58:03.45&   +61:05:57.6&  2.725$\pm$   1.217&  115.314$\pm$   12.254&    17.582&   1.302&   4231&    2.093    \\   
  55&  00:58:06.96&   +61:10:52.7&  2.412$\pm$   0.306&   87.118$\pm$    3.724&    15.030&   1.805&   3326&    1.336   \\   
  56&  00:58:08.18&   +61:11:23.6&  3.171$\pm$   0.132&   89.364$\pm$    1.210&    13.724&   1.880&   1918&    1.687   \\
\end{tabular}
\end{table} 
\startlongtable
\begin{table}
  \scriptsize
  \centering
      \begin{tabular}{llllllllr}\hline
          No.& RA (J2000)  & DEC (J2000)  & $P\pm\sigma_{P}$ & $\theta_{pos}\pm$ $\sigma_{\theta_{pos}}$&$V_{mag}$& $A_{V}$& $d$ & $P/A_{V}$   \\
  & (hh:mm:ss)  & (dd:mm:ss)   & (\%)  & ($^{\circ}$)  & (mag) & (mag) & (pc) & (\%/mag) \\
  \hline
  26&  00:59:09.71&   +60:51:14.3&  3.822$\pm$   0.486&   94.481$\pm$    3.573&    16.760&   1.676&   4970&    2.280  \\      
  27&  00:59:10.24&   +60:57:07.1&  2.119$\pm$   0.167&   89.030$\pm$    2.176&    13.733&   0.703&    947&    3.014  \\      
  28&  00:59:10.34&   +60:55:26.7&  3.448$\pm$   0.504&   90.528$\pm$    4.070&    16.199&   1.514&   2292&    2.277  \\      
  29&  00:59:11.15&   +60:51:48.2&  2.774$\pm$   0.735&   90.734$\pm$    7.378&    16.422&   1.489&   1007&    1.863  \\      
  30&  00:59:13.57&   +61:00:33.1&  2.213$\pm$   0.146&   76.950$\pm$    1.825&    14.266&   2.820&   1670&    0.785 \\      
  31&  00:59:15.41&   +60:51:41.5&  4.946$\pm$   0.958&   95.312$\pm$    5.442&    16.870&   1.298&   2232&    3.810  \\
\end{tabular}
\end{table} 
\startlongtable
\begin{table}
  \scriptsize
  \centering
      \begin{tabular}{llllllllr}\hline
          No.& RA (J2000)  & DEC (J2000)  & $P\pm\sigma_{P}$ & $\theta_{pos}\pm$ $\sigma_{\theta_{pos}}$&$V_{mag}$& $A_{V}$& $d$ & $P/A_{V}$   \\
  & (hh:mm:ss)  & (dd:mm:ss)   & (\%)  & ($^{\circ}$)  & (mag) & (mag) & (pc) & (\%/mag) \\
  \hline
  79&   01:00:13.98&   +60:51:47.3&  2.367$\pm$   0.684&  105.824$\pm$    8.013&    16.362&   1.082&   1650&    2.188   \\   
   80&   01:00:14.21&   +60:56:05.5&  2.451$\pm$   0.222&   88.042$\pm$    2.506&    15.047&   1.331&   3868&    1.841   \\   
   81&   01:00:14.22&   +60:59:43.0&  2.647$\pm$   0.340&   80.729$\pm$    3.555&    16.368&   1.090&   1935&    2.428   \\   
   82&   01:00:14.26&   +60:58:43.7&  4.000$\pm$   0.407&   91.987$\pm$    2.866&    16.713&   1.884&   2022&    2.123   \\   
   83&   01:00:14.60&   +60:53:58.0&  3.268$\pm$   0.268&   90.165$\pm$    2.275&    14.348&   1.360&   1637&    2.403    \\   
   84&   01:00:15.37&   +60:51:29.9&  2.595$\pm$   0.672&   91.743$\pm$    7.104&    16.781&   0.961&    755&    2.700   \\
\hline
  \end{tabular}
\end{table}

\end{document}